\documentclass{vgtc}                          

\ifpdf
  \pdfoutput=1\relax                   
  \usepackage{graphicx}                
  \DeclareGraphicsExtensions{.pdf,.jpg,.jpeg} 
\else
  \usepackage{graphicx}                
  \DeclareGraphicsExtensions{.eps}     
\fi%

\graphicspath{{jpg/}{npictures/}} 

\usepackage{microtype}                 
\PassOptionsToPackage{warn}{textcomp}  
\usepackage{textcomp}                  
\usepackage{mathptmx}                  
\usepackage{times}                     
\usepackage{cite}                      
\usepackage{tabu}                      
\usepackage{booktabs}                  
\usepackage{color}
\usepackage{amssymb}
\usepackage{stmaryrd}
\usepackage[]{algorithm2e}
\usepackage{amsmath}
\usepackage[noconfigs,american]{babel}

\usepackage{tikz}
\usetikzlibrary{fit,calc}

\colorlet{pink}{red!40}
\colorlet{blueoriginal}{blue!100}
\colorlet{blue}{cyan!60}
\colorlet{bluetwo}{cyan!10}

\makeatletter
\tikzset{%
  remember picture with id/.style={%
    remember picture,
    overlay,
    save picture id=#1,
  },
  save picture id/.code={%
    \edef\pgf@temp{#1}%
    \immediate\write\pgfutil@auxout{%
      \noexpand\savepointas{\pgf@temp}{\pgfpictureid}}%
  },
  if picture id/.code args={#1#2#3}{%
    \@ifundefined{save@pt@#1}{%
      \pgfkeysalso{#3}%
    }{
      \pgfkeysalso{#2}%
    }
  }
}

\def\savepointas#1#2{%
  \expandafter\gdef\csname save@pt@#1\endcsname{#2}%
}

\def\tmk@labeldef#1,#2\@nil{%
  \def\tmk@label{#1}%
  \def\tmk@def{#2}%
}

\tikzdeclarecoordinatesystem{pic}{%
  \pgfutil@in@,{#1}%
  \ifpgfutil@in@%
    \tmk@labeldef#1\@nil
  \else
    \tmk@labeldef#1,(0pt,0pt)\@nil
  \fi
  \@ifundefined{save@pt@\tmk@label}{%
    \tikz@scan@one@point\pgfutil@firstofone\tmk@def
  }{%
  \pgfsys@getposition{\csname save@pt@\tmk@label\endcsname}\save@orig@pic%
  \pgfsys@getposition{\pgfpictureid}\save@this@pic%
  \pgf@process{\pgfpointorigin\save@this@pic}%
  \pgf@xa=\pgf@x
  \pgf@ya=\pgf@y
  \pgf@process{\pgfpointorigin\save@orig@pic}%
  \advance\pgf@x by -\pgf@xa
  \advance\pgf@y by -\pgf@ya
  }%
}
\newcommand\tikzmark[2][]{%
\tikz[remember picture with id=#2] #1;}
\makeatother

\newcommand\MyBox[5][-1ex]{%
  \tikz[remember picture,overlay,pin distance=0cm]
  {\draw[draw=#4,line width=1pt,fill=#4!20,rectangle,rounded corners]
( $ (pic cs:#2) + (-1ex,2ex) $ ) rectangle ( $ (pic cs:#3) + (#5,#1) $ );
}
}

\newcommand*{\blueish}[1]{%
  \tikz[baseline=(X.base)]
  \node[rectangle, draw=blue, fill=bluetwo, rounded corners, inner sep=0.0mm, line width=1pt] (X) {
  #1
  };%
}

\onlineid{1020}

\vgtccategory{Research}

\vgtcinsertpkg

\title{Lifted Wasserstein Matcher for Fast and Robust Topology Tracking}

\newcommand{\maxime}[1]{#1}

\author{\maxime{Maxime Soler}\thanks{\maxime{e-mail: soler.maxime@total.com}}\\ %
        \parbox{1.4in}{\scriptsize   
        \centering \maxime{Total S.A. \\ Sorbonne Universit\'e, CNRS \\
        Laboratoire d'Informatique de Paris 6, F-75005 Paris}}
\and \maxime{M\'elanie Plainchault}\thanks{\maxime{e-mail: melanie.plainchault@total.com}}\\ %
     \scriptsize \maxime{Total S.A.} %
\and \maxime{Bruno Conche}\thanks{\maxime{e-mail: bruno.conche@total.com}}\\ %
     \scriptsize \maxime{Total S.A.} %
\and \maxime{Julien Tierny}\thanks{\maxime{e-mail: julien.tierny@sorbonne-universite.fr}}\\ %
     \parbox{1.4in}{\scriptsize \centering \maxime{Sorbonne Universit\'e, CNRS \\
     Laboratoire d'Informatique de Paris 6, F-75005 Paris}}
     }

\newcommand{\removed}[1]{}

\newcommand{\eqrref}[1]{Eq.~\ref{#1}}
\newcommand{\figref}[1]{Fig.~\ref{#1}}
\newcommand{\thref}[1]{Theorem~\ref{#1}}
\newcommand{\secref}[1]{Sec.~\ref{#1}}
\newcommand{\tabref}[1]{Table~\ref{#1}}
\newcommand{\algref}[1]{Algorithm~\ref{#1}}

\newcommand{\domain}{\mathcal{M}}
\newcommand{\range}{\mathbb{R}}

\newcommand{\sub}[1]{f^{-1}_{-\infty}(#1)}
\newcommand{\sur}[1]{f^{-1}_{+\infty}(#1)}

\newcommand{\persistentDiagram}[1]{\mathcal{D}(#1)}
\newcommand{\Index}{\mathcal{I}}

\newcommand{\wasserstein}{d^W_\nu}

\abstract{
This paper presents a robust and efficient method for tracking topological
features in time-varying scalar data. Structures are tracked based on the
optimal matching between persistence diagrams with respect to the
 Wasserstein metric. This fundamentally relies on solving the assignment
problem, a special case of optimal transport, for all consecutive
timesteps. Our approach relies on two main contributions. First, we
revisit the seminal assignment algorithm by Kuhn and Munkres which we
specifically adapt to the problem of matching persistence diagrams
in an efficient way. Second, we propose an extension of the Wasserstein
metric that significantly improves the geometrical stability of the
matching of domain-embedded persistence pairs. We show that this
geometrical lifting has the additional positive side-effect of improving
the assignment matrix sparsity and therefore computing time. 
The global framework 
computes persistence
diagrams and finds optimal matchings in parallel for every 
consecutive timestep. Critical trajectories are constructed by associating
successively matched persistence pairs over time. Merging and splitting
events are detected with a geometrical threshold in a post-processing
stage. Extensive experiments on real-life datasets show that our matching
approach is up to two orders of magnitude faster than the seminal Munkres
algorithm. Moreover, compared to a modern approximation method,
our approach provides competitive runtimes while guaranteeing exact results.
We demonstrate the utility of our global framework by extracting critical
point trajectories from various\removed{simulated} time-varying datasets and
compare it to the existing methods based on associated overlaps of
volumes. Robustness to noise and temporal resolution downsampling is
empirically demonstrated.
} 


\begin{document}


\firstsection{Introduction}

\maketitle

Performing feature extraction and object tracking
is an important topic in scientific visualization,
for it is key to understanding
time-varying data.
Specifically, it allows to detect and track the evolution of regions
of interest over time, which is central to
many scientific domains, such as
combustion \cite{bremer_tvcg11},
aerodynamics \cite{Hannon94},
oceanography \cite{ringler13} or meteorology \cite{FZhang07}.
With the increasing power of computational resources
and resolution of acquiring devices, 
efficient methods are needed to enable the analysis 
of large 
datasets.

The emergence of new paradigms for scientific simulation,
such as \emph{in-situ} and \emph{in-transit}
\cite{bennett12, yu10, Rivi2011InsituVS, Rasquin11, Moreland2011},
clearly exhibits
the ambition to
reach toward exascale computing \cite{chris13} in the forthcoming years.
In this context,
as both spatial and temporal resolutions of acquired or simulated
datasets keep on increasing,
understanding
the evolution of features of interest
throughout time proves challenging.

Topological data analysis has been used
in the last decades as a
robust and reliable setting
for hierarchically defining features
in scalar data \cite{edelsbrunner09}.
In particular, its successful
application to time-varying data \cite{bajaj06, bremer10}
\maxime{makes} it a prime candidate for tracking.
Both topological analysis and feature tracking
have been applied \emph{in-situ}
\cite{zhang12, landge14}, which demonstrates
their interest in the context of large-scale data.
Nonetheless, major bottlenecks of state-of-the art
topology tracking methods
are still the high required computation cost 
as well as the need for high temporal resolution. 

In this paper, we propose a novel feature-tracking
framework, which correlates topological features
in time-varying data in an efficient and meaningful way.
It is the first approach, to the best of our knowledge,
combining the setting of topological data analysis
with optimal transport for the problem of feature tracking.
More precisely, the key idea is to use
combinatorial optimization for matching
topological structures (namely, \emph{persistence diagrams})
according to a fine-tuned metric.
After exposing our formal setting (\secref{sec:preliminaries}),
we introduce an extension of the exact assignment algorithm by Kuhn and Munkres
\cite{Kuhn55thehungarian, Munkres57algorithmsfor} that we adapt
in an efficient way to the case of persistence diagrams
(\secref{sec:persistenceCriterion}).
We highlight the issues raised by the classical Wasserstein
metric between diagrams,
and propose a robust \emph{lifted} metric that overcomes these limitations
(\secref{sec:liftedPersistenceMatching}).
We then present the detailed tracking
framework (\secref{sec:featureTracking}).
Extensive experiments 
demonstrate the \maxime{utility} of our approach
(\secref{sec:results}).

\subsection{Related work}

Our framework encompasses the definition,
correlation and tracking
of topological features in scalar fields. As such, it is
related to topological data \maxime{analysis} of scalar fields,
tracking techniques, the definition of metrics and
combinatorial optimization.

\noindent
\textbf{Topological analysis techniques} \cite{edelsbrunner09,
pascucci_topoInVis10, heine16, Defl15} have demonstrated their
ability over recent years to
capture \maxime{features of interest in scalar data} in a generic, robust
\cite{edelsbrunner02, cohen-steiner05} and efficient
manner, for many applications as
turbulent combustion \cite{bremer_tvcg11},
computational fluid dynamics \cite{favelier16},
material sciences \cite{gyulassy_vis15},
chemistry \cite{chemistry_vis14}, astrophysics \cite{sousbie11},
medical imaging \cite{carr04, bock18}.
One reason for their success
in applications is the possibility for domain experts to
easily translate high level \maxime{structural} notions
\maxime{into} topological abstractions,
such as contour trees \cite{carr2003computing}, Reeb Graphs
\cite{pascucci07, biasotti08},
 Morse-Smale complexes \cite{gyulassy_vis08}.
For instance, \maxime{in astrophysics} the cosmic web can be extracted by querying the
most persistent \maxime{1-}separatrices of the Morse-Smale complex
connected to maxima of matter density \cite{sousbie11}.
Similar domain-specific notions are translated
into topological terms in the above examples.

\noindent
\textbf{Feature tracking:}
Topology has been used for feature extraction and tracking in the
context of vector fields
\cite{tricoche02, post03, reininghaus2012efficient},
mostly relying on
stream lines, path lines
\cite{theisel03, theisel04, theisel05, shi06a, Shi2009},
or tracking \maxime{punctual} singularities \cite{klein07}.
For the latter, a forward streamline integration of
critical points is performed
in a specific scale space,
which adapted for time-varying data would
require knowledge about the evolution of the field,
and for instance to compute time-derivatives.

\maxime{F}or scalar data, features are defined based on attributes that are either
geometric (isosurfaces, thresholded regions),
or topological (contour trees, Reeb graphs)\maxime{. Similarly,}
tracking approaches either rely on geometrical
(volume overlaps, distance between centers of gravity) or
topological extracts (Jacobi set, segment overlap).

Geometrical approaches are based on
thresholded connected components \cite{silver99},
glyphs \cite{Reinders2001},
cluster tracking \cite{Grottel07},
petri nets \cite{Ozer14},
or propose a hierarchical representation \cite{Gu11}.
Similarly,
the core methodology for associating topological
features for tracking is often based on overlaps of geometrical domains
along with other attributes
\cite{Saikia17,
bremer10, bremer_tvcg11, bajaj06, Silver95, Silver97, Silver972, Silver98},
on tracking Jacobi sets \cite{edelsbrunnerJacobi, edelsbrunner04},
or matching isosurfaces in higher-dimensional spaces \cite{Ji04, Ji03}.

Such approaches usually test features in two consecutive timesteps
against one another for potential overlaps,
then draw the best correspondence between features
according to some criterion. Typical criteria include
\maxime{optimizing} the overlapping volume, mass, distance between centroids,
or a combination of these \cite{Samtaney94, Reinders2001}.
For this to work,\removed{it is required that}
the temporal sampling rate of the underlying data \maxime{must} be such
that features in two consecutive timesteps
\maxime{effectively}\removed{do} overlap.
This first criterion is thus not very robust to
temporal downsampling.

Other approaches rely on global optimization \cite{Ji2006},
using the Earth Mover's distance \cite{levina01}
between geometrical features with various attributes such as
centroid position, volume and mass.
This does not\maxime{,} however\maxime{,} benefit from the natural
definition of features offered by topological data analysis,
nor from the possibility
to simplify features in a hierarchical way.
This is a real drawback in the context of noisy data as it
implies dealing with large,
computation-intensive optimization problems
between every pair of timesteps.

Once features have been defined, and a methodology established
to associate them in consecutive timesteps,
the tracking representation is quite independent
of whether geometrical or topological arguments have been used.
Most often, graphs are used \cite{robbins00, wida15, laney_vis06}, such as
Reeb graphs \cite{Weber2011, edelsbrunner04} and
nested tracking graphs \cite{Lukasczyk17}. Many popular graph
structures are accounted for in \cite{Wang17}.
An inconvenience of extracting rich
tracking structures such as these without taking careful
attention to potential noise is that it makes the
interpretation quite difficult.
In \cite{bremer_tvcg11}, the tracking graph is dense and
intricate, making exploration impractical.
It is therefore mandatory to do filtering and simplification
in a post-processing stage,
or to cleverly discard noisy events beforehand.

\noindent
\textbf{Assignment problem\maxime{s}:}
\maxime{S}ince we revisit the original algorithm by Kuhn and Munkres, we discuss here
some related work in combinatorial optimization.
The assignment problem is the discrete optimization problem
consisting of finding a perfect matching
of optimal cost in a weighted bipartite graph
\cite{Munkres57algorithmsfor, Bertsekas98, Burkard09}.
In other terms, the problem \maxime{is to find} an optimal one-to-one
correspondence between discrete entities
(such as singularities in a scalar field at two \maxime{different} timesteps),
with a cost associated to each possible correspondence.
It can be solved with the seminal
Kuhn-Munkres algorithm \cite{Munkres57algorithmsfor}.
The auction algorithm \cite{Bertsekas98, bertsekas89, kerber17}
is another popular approach for solving the assignment problem with a
user-defined error threshold on the resulting assignment cost.
In practice, this threshold is often set to 1\% of the scalar range.
A more general, continuous formulation of this problem is at the heart
of Transportation theory \cite{monge81, Kantorovich42, Villani08}.
Modern techniques \cite{Cuturi13} have attracted acute interest
for making this theoretical setup central to
shape correlation \cite{solomon16} and interpolation \cite{solomon17},
which do bear resemblance to feature tracking.

\noindent
\textbf{Metrics:}
\maxime{S}ince we introduce a new metric for the matching of persistence diagrams, we
discuss in the following existing metrics traditionally used in topological
data analysis. The Bottleneck and the interleaving distances
\cite{cohen-steiner05, ChazalCGGO09} have been widely investigated
to study the stability of persistence diagrams. 
\maxime{These metrics have been notably adapted in the context of
kernel methods \cite{ReininghausHBK15, CarriereCO17} in machine learning.} 
In particular, the
Bottleneck distance is a special case of the more general \emph{Wasserstein}
metric \cite{monge81, Kantorovich42}
applied to diagram points, also known as the
\emph{Earth Mover's Distance} \cite{levina01}.
The standard approach for computing the discrete Wasserstein metric
relies on solving the associated assignment problem,
either with an exact Kuhn-Munkres approach \cite{dionysus, weaver13}
or with an auction-based approximation \cite{kerber17}.
However, as discussed in
\secref{sec:liftedPersistenceMatching}, when
\maxime{these methods (metric-based \cite{ChazalCGGO09, cohen-steiner05} or
kernel-based \cite{ReininghausHBK15, CarriereCO17}) are}
applied as-is for tracking
purposes, a high geometrical instability occurs which impairs the tracking
robustness, as already observed in the case of
\emph{vineyards} \cite{Cohen-Steiner06}.
Our work (see \secref{sec:liftedPersistenceMatching}) addresses this issue.

\subsection{Contributions}
This paper makes the following new contributions:
\begin{enumerate}
  \vspace{-1ex}
  \item{\textbf{Approach:} We present a sound and original framework,
  which is the first
  combining topology and transportation for feature tracking,
  comparing \maxime{favorably} to other state-of-the-art approaches, both
  in terms of speed and robustness.}
  \vspace{-1ex}
  \item{\textbf{Metric:} We extend \maxime{traditional} topological metrics,
  for the needs of time-varying feature tracking, notably enhancing
  geometrical stability
  and computing time.
  }
  \vspace{-1ex}
  \item{\textbf{Algorithm:} We 
  extend the assignment
  method by Kuhn and Munkres to solve the problem of persistence matchings
  in a fast and exact way, taking advantage of our metric.
  }
\end{enumerate}


\section{Preliminaries}
\label{sec:preliminaries}
This section describes our formal setting and presents an overview of
our approach. 
It contains definitions that we adapted from Tierny et al. \cite{ttk}.
An introduction to topology can be found in
\cite{edelsbrunner09}.

\subsection{Background}
\label{sec:background}

\noindent
\textbf{Input data:}
Without loss of generality, we assume that the input data is a piecewise
linear (PL) scalar field $f : \domain \rightarrow \mathbb{R}$ defined on a
PL $d$-manifold $\domain$ with $d \le 3$.
Values are given at the vertices of $\domain$ and linearly
interpolated on higher dimension simplices.

\noindent
\textbf{Critical points:}
Given an isovalue $i \in \range$,
the \emph{sub-level set} of $i$,
noted $\sub{i}$, is the pre-image of the open
interval $(-\infty, i)$ onto $\domain$ through $f$:
$\sub{i} = \{ p \in \mathcal{M} ~ | ~ f(p) < i \}$. The
\emph{sur-level set} is \maxime{symmetrically} given by $\sur{i} = \{ p \in \mathcal{M} ~
| ~ f(p) > i \}$. These two objects serve as
segmentation tools in many analysis tasks \cite{bremer_tvcg11}.

The points $p \in \domain$
where the topology
of $\sub{f(p)-\epsilon}$
\maxime{differs} from
that
of $\sub{f(p)+\epsilon}$
are the \emph{critical points} of $f$
and their values are called
\emph{critical values}.
Critical points can be classified with their \emph{index} $\Index$, which
is 0 for minima, 
1 for 1-saddles, 
$d - 1$ for $(d-1)$-saddles 
and $d$ for maxima, 
with $d$ the dimension of $\domain$.

\begin{figure}[tb]
 \centering
 \includegraphics[width=\columnwidth]{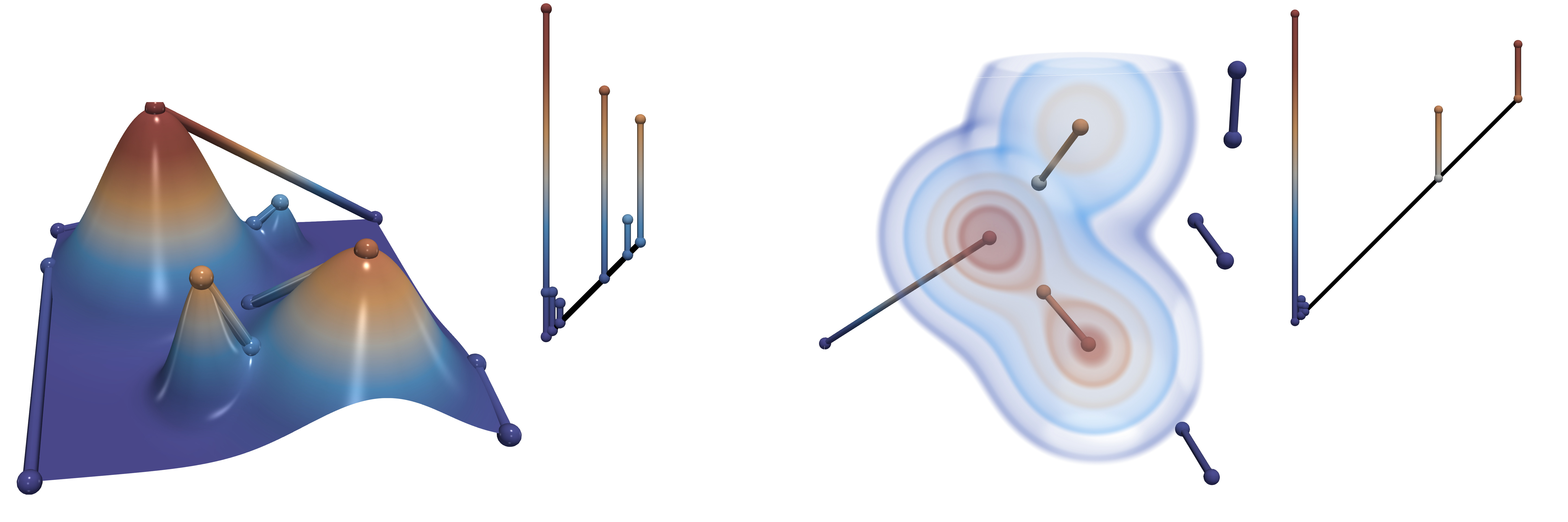}
 \caption{Four gaussians defined on a 2D plane (top left) and
 the associated persistence diagram (right);
 three gaussians defined on a 3D volume (bottom left) with
 persistence diagram (right). The most prominent topological features
 are those which have the longest \emph{lifespan}, and
 correspond to the longest vertical bars in persistence diagrams.
 }
 \label{fig:persistence2D}
 \vspace{-2ex}
\end{figure}

\noindent
\textbf{Persistence diagrams:}
The distribution of critical points of $f$ can \maxime{visually} be
represented by a topological abstraction called the \emph{persistence
diagram} \cite{edelsbrunner02, cohen-steiner05} (\figref{fig:persistence2D}).
By applying the Elder Rule \cite{edelsbrunner09}, critical points can be
arranged in a set of pairs, such that each critical point appears in only one
pair $(c_i, c_j)$ with $f(c_i) < f(c_j)$ and $\Index(c_i) = \Index(c_j) - 1$.
More precisely, the Elder Rule states that as the value $i$ increases, if two
topological features of $\sub{i}$, for instance two connected components, meet
at a given saddle $c_j$ of $f$, the \emph{youngest} of the two 
(the
one with the highest minimal value, $c_i$) \emph{dies} at the advantage of the oldest
(the one with the lowest minimal value). Critical points $c_i$ and $c_j$ then
form a \emph{persistence pair}.


The persistence diagram $\persistentDiagram{f}$ embeds each pair $(c_i, c_j)$
in the plane such that its horizontal coordinate equals $f(c_i)$, and the vertical coordinate of
both $c_i$ and $c_j$ is $f(c_i)$ and $f(c_j)$, corresponding respectively
to the \emph{birth} and \emph{death} of the pair. The height of the pair
$P(c_i, c_j) = |f(c_j) - f(c_i)|$ is called the \emph{persistence} and denotes the
life-span of the topological feature created at $c_i$ and destroyed at $c_j$.
In three dimensions, the persistence of the pairs linking critical points of index
$(0,1)$, $(2, 3)$ and $(1,2)$ denotes the life-span of
connected components, voids and non-collapsible cycles of $\sub{i}$.

In practice, persistence diagrams serve as an important visual representation
of the distribution of critical points in a scalar data-set.
Small oscillations in the data due to noise are typically represented by
critical point pairs with low persistence, in the vicinity of the diagonal.
In contrast, topological features that are the most
prominent in the data are associated with large vertical bars
(\figref{fig:persistence2D}). In many
applications, persistence diagrams help users as a visual guide to
interactively tune simplification thresholds in 
\removed{topology-based,} multi-scale
data segmentation tasks based on the Reeb graph \cite{reeb46, carr04,
pascucci07, tierny_vis09, gueunet_ldav16,
tierny_vis16}
or the Morse-Smale complex \cite{gyulassy_vis08, gyulassy_vis14, robins11}.

\noindent
\textbf{Wasserstein distance:}
Metrics have been defined to evaluate the distance between
two scalar fields $f, g : \domain \rightarrow \range$.
The $L$-norm
$||f - g||_\nu$, is a classical example.
In the context of topological data analysis, multiple metrics
\cite{cohen-steiner05, ChazalCGGO09} have been introduced in order to
compare persistence diagrams. In our context, such metrics are
key to identifying zones in the data which are similar to one another.
Persistence diagrams can be associated with a pointwise distance, noted
$d_\nu$ inspired by the $L$-norm.
Given persistence pairs $a = (a_x, a_y) \in \persistentDiagram{f}$ and $b = (b_x, b_y) \in
\persistentDiagram{g}$, $d_\nu$ is given by \eqrref{eq:dnu}:
\vspace{-1ex}
\begin{equation}
\begin{split}
  d_\nu (a, b) &= (|a_x - b_x|^\nu + |a_y - b_y|^\nu)^{1/\nu} \\
\end{split}
  \label{eq:dnu}
\end{equation}
\vspace{-3ex}

\noindent
The \emph{Wasserstein} distance \cite{monge81, Kantorovich42},
sometimes called the \emph{Earth Mover's Distance} \cite{levina01},
noted $\wasserstein$, between the persistence diagrams
$\persistentDiagram{f}$ and $\persistentDiagram{g}$ is defined in \eqrref{eq:wass}:
\vspace{-1.5ex}
\begin{eqnarray}
  \wasserstein \big(\persistentDiagram{f}, \persistentDiagram{g}\big) =
    \min_{\phi \in \Phi}
      \bigg(
       \sum_{a \in \persistentDiagram{f}}
         d_\nu
          \big(
            a, \phi(a)
          \big)^\nu
      \bigg)^{1/\nu}
      \label{eq:wass}
\end{eqnarray}
\vspace{-2ex}

\noindent
where $\Phi$ is the set of all possible bijections $\phi$ mapping
each critical point $a$
of $\persistentDiagram{f}$ to
a critical point $b$ of the same index
$\Index$ in $\persistentDiagram{g}$ or to
the diagonal, noted $\maxime{\text{diag}}(a)$
\maxime{--} which corresponds to the removal of the corresponding feature
from the assignment, with \maxime{a} cost $d(a,\maxime{\text{diag}}(a))$.

\subsection{Assignment problem}
\label{sec:assignmentOverview}


The assignment problem is the problem of choosing an optimal assignment
of $n$ workers $w\in W$ to $n$ jobs $j\in J$, assuming numerical ratings are given for each
worker's performance on each job.

Given ratings $r(w_x, j_y)$ are
summed up in a cost matrix $(r_{xy})$,
finding an optimal assignment means
choosing a set of $n$ \emph{independent}
entries 
of the matrix so that the sum
of these elements is optimal.
\emph{Independent} means than no two such elements
should belong to the same row or column (i.e.
no two workers should be assigned to the same job and no worker should be 
given more than one job).
In other words, one must find a map $\sigma:W\rightarrow J$
of workers and jobs for which the sum $\sum_x(r(w_x,\sigma(w_x)))$ is optimal.
There are $n!$ possible assignments, of which several may be optimal, so that an
exhaustive search is impractical as $n$ gets large.

Similarly, the unbalanced assignment problem is the problem of finding an optimal
assignment of $n$ workers to $m$ jobs, where some jobs or workers might be left
unassigned. This is the case of assignments between sets of persistence pairs;
where costs are defined for leaving specific pairs unassigned.

The Hungarian algorithm \cite{Kuhn55thehungarian, Munkres57algorithmsfor} is the first polynomial
algorithm proposed by Kuhn to solve the assignment problem.
It is an iterative algorithm based on the following two properties:

\newtheorem{propagation}{Theorem}
\vspace{-1.5ex}
\begin{propagation}
 If a number is added or subtracted from all the entries of any one row or column
 of a cost matrix, then an optimal assignment for the resulting cost matrix is also
 an optimal assignment for the original cost matrix.
 \label{th:propagation}
 \vspace{-2ex}
\end{propagation}

This means that the cost matrix $(r_{ij})$ can be replaced
with $(r_{ij})-u_i-v_j$ where $u_i$ (resp. $v_j$) is an arbitrary number
which is fixed for the $i^{th}$ row (resp. the $j^{th}$ column).

\newtheorem{stopCriterion}[propagation]{Theorem}
 \vspace{-1.5ex}
\begin{stopCriterion}
 If $R$ is a matrix and $m$ is the maximum number of independent
 zeros of $R$ (i.e. number of entries valued at 0),
 then there are $m$ lines (row or columns)
 which contain all the zeros of $R$.
 \label{th:stop}
 \vspace{-1.5ex}
\end{stopCriterion}
This allows to determine whether an optimal assignment has been found
and thus constitutes the stop criterion.

The algorithm iteratively performs additions and subtractions
on lines and columns of the cost matrix, in a way that globally decreases the
matrix cost, until the optimal assignment has been found, that is, until
the matrix contains a set of $min(m,n)$ independent zeros.

In the remainder we consider the $O(min(m,n)^2 max(m,n))$ unbalanced Kuhn-Munkres
algorithm \cite{Munkres57algorithmsfor, Bourgeois71}, an improvement over Kuhn's original
version which follows the same principles, with an enhanced strategy for
finding independent elements. The goal is always to reduce the cost matrix
and find a maximal set of independent zeros.
These independent zeros are marked with a \emph{star}:
they are candidates for the optimal assignment.
Zeros which are candidates for being swapped with a starred zero are marked with a \emph{prime}.
Throughout the algorithm, rows and columns of the matrix
are marked as \emph{covered} to restrict the search for 
\maxime{candidate zeros.}

\begin{figure}[tb]
 \centering
 \includegraphics[width=\columnwidth]{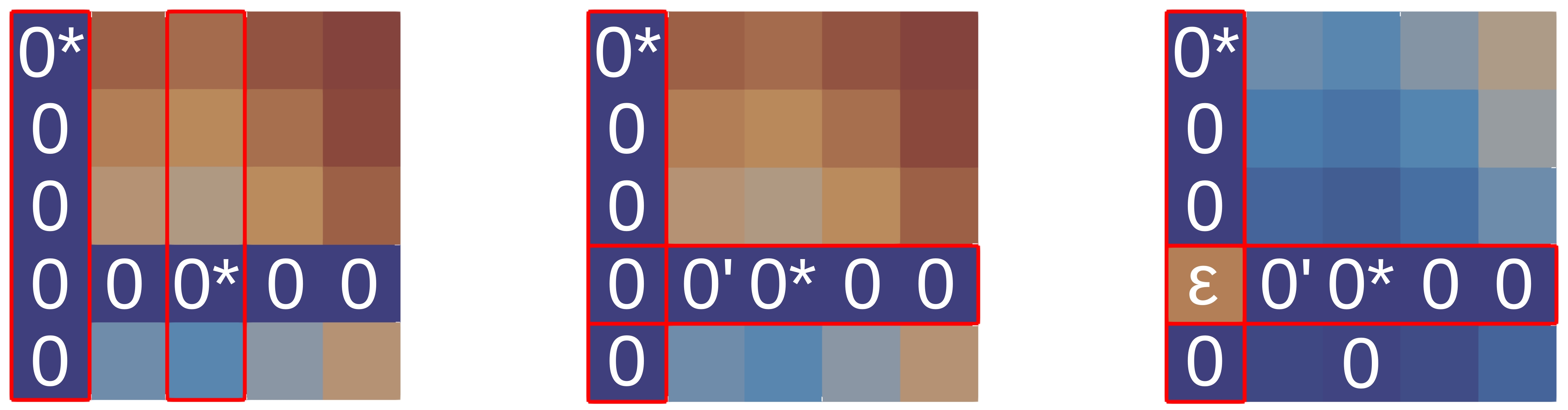}
 \caption{Matrix reduction phase.
 Subtracting the minimum element from each of the $n$ rows and columns might
 not be sufficient to make a set of $n$ \emph{independent} zeros appear.
 In the above example, initially detected independent zeros are first starred.
 All columns containing a 0* are then covered (left).
 An uncovered zero which has a 0* in its row is found and primed;
 its row is covered and the column of the 0* is uncovered (center).
 At this point, all zeros are covered by construction.
 Let $\epsilon$ be the smallest uncovered value.
 Add $\epsilon$ to every covered row;
 subtract $\epsilon$ from every uncovered column.
 This amounts to decreasing uncovered elements by $\epsilon$
 and increasing twice-covered elements by $\epsilon$.
 The sum of the elements of the matrix has been decreased and a new
 zero has appeared in an uncovered zone.
 }
 \label{fig:reduce}
 \vspace{-2ex}
\end{figure}

\begin{figure}[tb]
 \centering
 \includegraphics[width=\columnwidth]{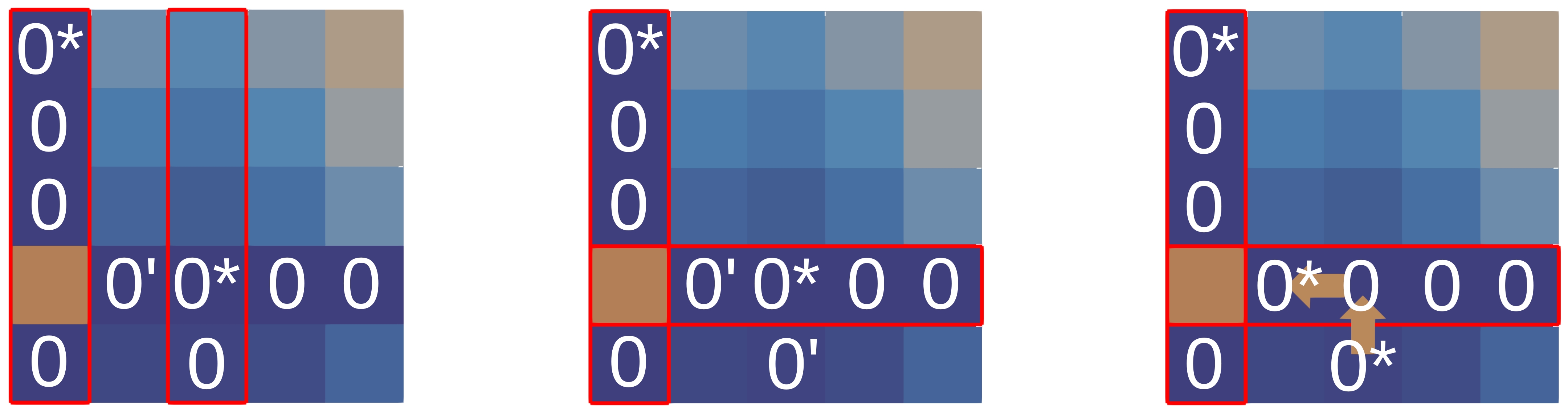}
 \caption{
 Augmenting path phase. After the first non-covered zero (left) is primed
 and covers updated, there \maxime{is} one non-covered zero $Z_1$
 in the matrix (center), which is then primed.
 Let $Z_2$ be the 0* in the column of $Z_1$ (if any),
 let $Z_3$ be the 0' in the row of $Z_2$ (there is one).
 Consider the series consisting of \maxime{0*} ($Z_{2i}$) and \maxime{0'}
 ($Z_{2i+1}$) until it ends at a 0' that has no 0* in its column.
 Unstar each 0*\maxime{,} star each 0' of the series.
 The number of starred zeros has increased by one.
 }
 \label{fig:augment}
 \vspace{-2ex}
\end{figure}

The algorithm can be seen as two alternating phases:
a matrix reduction phase (\figref{fig:reduce}) which makes new zeros appear,
and an augmenting path phase (\figref{fig:augment}) which augments the number of
marked (\emph{starred}) independent zeros.

\subsection{Persistence assignment problem}
\label{sec:persistenceMatching}
The assignment problem for persistence pairs is very similar to the standard
unbalanced assignment problem, except additional costs are defined
for not assigning elements (\emph{i.e.} matching persistence pairs with the diagonal).
The assignment between diagrams $P$ and $Q$ then involves $r_{ij}$
the numerical rating associated with assigning $p_i \in P$ with $q_j \in Q$,
along with $r_{i,-1}$ (resp. $r_{-1,j}$) the numerical rating associated with
matching $p_i$ (resp. $q_j$) with the diagonal.

If $P$ and $Q$ are sets of persistence pairs such that $card(P)=n$ and $card(Q)=m$,
then it is possible to solve the persistence assignment problem using
the standard Kuhn-Munkres algorithm with the $(n+m)\times(n+m)$ cost matrix
described by \eqrref{eq:bigmatrix},
as proposed in \cite{dionysus}:

\vspace{-3.5ex}
\begin{equation}
 r_{ij}=\begin{cases}
    d_{\nu}(p_i, q_j)    & {\scriptstyle\text{if  $0 < i \le n, 0 < j \le m$}}\\
    d_{\nu}(q_{j}, \maxime{\text{diag}}(q_{j}))        & {\scriptstyle\text{if  $n < i \le m+n, 0 < j \le m$}}\\
    d_{\nu}(p_{i}, \maxime{\text{diag}}(p_{i}))        & {\scriptstyle\text{if  $0 < i \le n, m < j \le m+n$}}\\
    0                    & {\scriptstyle\text{if  $n < i \le m+n, m < j \le n+m$}}
  \end{cases}
 \label{eq:bigmatrix}
\end{equation}

The first line corresponds to matching pairs from $P$ to pairs from $Q$;
the second one corresponds to the possibility of matching pairs from
$P$ to 
the diagonal; the third one is for matching pairs of $Q$ to 
the diagonal and
the last one completes the cost matrix.
The drawback of this approach is that it requires to solve the assignment problem
on a $(n+m)^2$ cost matrix (that potentially contains two non-sparse blocks
where persistence elements are located, see \figref{fig:representations}),
though the number of distinct elements is at most $(n+1)\times (m+1)$.
As seen in \secref{sec:persistenceCriterion}, our algorithm addresses \maxime{this} issue.

\subsection{Overview}
This section presents a quick overview of our tracking method,
which is illustrated in \figref{fig:overview}.
The input data is a time-varying PL scalar field $f$ defined on a
PL $d$-manifold $\domain$ with $d \le 3$.

\begin{enumerate}
  \vspace{-1.5ex}
  \item{
  First, we compute the persistence diagram of the scalar field for every
  available timestep.
  }
  \vspace{-1.5ex}
  \item{
    Next, for each pair of two consecutive timesteps $t$ and $t+1$,
    we consider the two corresponding persistence diagrams
     $\persistentDiagram{f_{t}}$ and
    $\persistentDiagram{f_{t+1}}$.
    For each couple of persistence pairs
    $(p_i, q_j) \in \persistentDiagram{f_{t}} \times \persistentDiagram{f_{t+1}} $,
    we define a distance metric corresponding to the similarity of these pairs:
    $d_\nu(p_i, q_j)$
    (see \secref{sec:liftedPersistenceMatching}).
  }
  \vspace{-1.5ex}
  \item{
    For each pair of consecutive timesteps,
    we compute a \emph{matching function} $M$.
    Every persistence pair $p_i$ of $\persistentDiagram{f_{t}}$
    is associated to
    to $M(p_i)$, which is either a
    persistence pair $q_j$ in $\persistentDiagram{f_{t+1}}$
    or 
    $\maxime{\text{diag}}(p_i)$ 
    so as to minimize the total distance $\sum_{i}d(p_i, M(p_i))$.
    Finding the optimal
    $M$ involves solving a variant of the classical Assignment Problem,
    as presented in \secref{sec:persistenceMatching}.
    Only persistence pairs involving critical points of the same index are
    \maxime{taken} into account.
  }
  \vspace{-1.5ex}
  \item{
    We compute tracking trajectories starting from the first timestep.
    If at timestep $t$ the matching associates $p_i$ with $M_t(p_i) = q_j$,
    then a segment is traced between $p_i$ and $q_j$.
    If $M_t(p_i) = \maxime{\text{diag}}(p_i)$, the current trajectory ends.
    Trajectories are grown following this principle throughout all timesteps.
    Properties are associated to trajectories (time span\maxime{, critical index}), and to
    trajectory segments (matching cost, scalar value).
  }
  \vspace{-1.5ex}
  \item{
    Finally, trajectories are post-processed to detect feature
    merging or splitting events with a user-defined geometric threshold.
  }
\end{enumerate}


\section{Optimized persistence matching}
\label{sec:persistenceCriterion}
This section presents our novel extension of the Kuhn-Munkres algorithm,
which has been specifically designed to address the computation time 
bottleneck described in \secref{sec:assignmentOverview}.

\subsection{Reduced cost matrix}

\begin{figure}[h]
 \centering
 \includegraphics[width=\columnwidth]{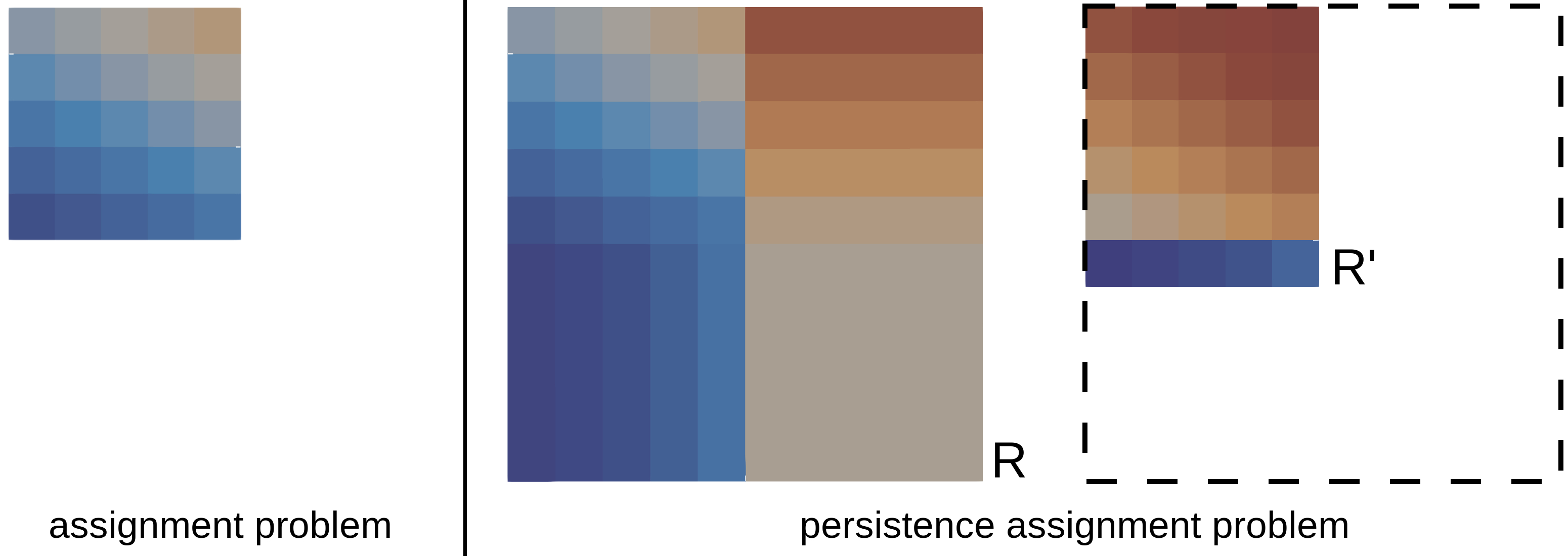}
 \caption{Cost matrices for a balanced assignment problem (left, $n\times n$ elements);
 for a persistence assignment problem with \cite{dionysus}
 (center, $R$ with $2n\times 2n$ elements -- \eqrref{eq:bigmatrix}); and for the same 
 persistence assignment problem with our proposed
 approach (right, $R'$ with $(n+1)\times n$ elements -- \eqrref{eq:smallmatrix}).
 Persistence elements in $R$ 
 induce two redundant
 non-sparse blocks (top-right and bottom-left).
 }
 \label{fig:representations}
 \vspace{-2ex}
\end{figure}

The classical persistence assignment algorithm based on Kuhn-Munkres considers $R$,
a $(n+m)^2$ cost matrix.
We propose to work instead with $R'$, a reduced $(n+1)\times m$ matrix
defined in \eqrref{eq:smallmatrix},
where every 
zero appearing
in the last row
is considered independent.
This amounts to considering
that persistence pairs corresponding to rows are not assigned by default.
\figref{fig:representations} summarizes the
matrices considered by each assignment method.
\vspace{-2ex}
\begin{equation}
r_{ij}'=\begin{cases}
    d_{\nu}(p_i, q_j) - \maxime{d_{\nu}(\text{diag}}(p_i),p_i) & {\scriptstyle\text{if  $0 < i \le n, 0 < j \le m$}}\\
    \maxime{d_{\nu}(\text{diag}}(q_j),q_j)        & {\scriptstyle\text{if  $i = n+1, 0 < j \le m$}}
  \end{cases}
 \label{eq:smallmatrix}
\end{equation}
\vspace{-2ex}

This last row, emulating the diagonal blocks of \figref{fig:representations}-b
requires a specific handling in the optimization procedure.
In particular, it 
requires the first step of the
algorithm to subtract minimum elements from columns (and not rows) so as not to
have negative elements in the matrix.

As a reminder,
the original algorithm proceeds iteratively in two alternating phases:
matrix reduction that makes new zeros appear,
and augmenting path that finds a maximal set of independent zeros.
At the $i^{th}$ iteration, the current maximal set of independent
zeros is made of \emph{starred} zeros. After a matrix reduction,
new zeros appeared that can potentially belong to the new maximal set of
independent zeros. Such candidates are \emph{primed}.
A single augmenting path (as in \figref{fig:augment})
replaces a set of $n$ \emph{starred} zeros
with $n+1$ \emph{primed} zeros, forming a new set of independent zeros
with one more element.
Rows and columns of the matrix are
marked as
\emph{covered} to restrict the
search for candidates in the augmenting path phase.
Blue blocks of \algref{alg:sparse} indicate our extension of 
the Kuhn-Munkres algorithm.

In this novel extension, 
an augmenting path constructed in the corresponding phase can start from
a starred zero in the last row (and then potentially find a primed zero
in its column), but such a path can never access a starred zero in the
last row at another step, for the corresponding column would have been
covered prior to this (and thus cannot contain a primed zero,
see \algref{alg:sparse}).
A starred zero in the last row can then never be unstarred.

The Kuhn-Munkres approach has the property to only increase
row values (and only decrease column values). When our algorithm
working on the reduced matrix $R'$ ends,
it is therefore not possible that the elements on the \maxime{top-right} corner of the
corresponding full matrix $R$ be negative.
Furthermore, given \thref{th:propagation},
the resulting matrix corresponds to the same assignment problem.

\subsection{Optimality}
\label{sec:optimality}

Working with the reduced matrix $R'$, however, does not necessarily yield an
optimal assignment. When assignments are found in the bottom row, if
there has been additions to the matrix rows, then the corresponding $R$ matrix
would contain a top-right block that is not zero, and a top-left block that is
not zero either. Thus, the stop criterion stated by \thref{th:stop} may not
be respected when $k=\min(m, n)$ lines are covered (as the real number of
independent zeros in $R$ is $m+n$).
Moreover, in our setup, a starred zero in the last column can never
be unstarred; this is allowed in the approach on $R$,
owing to the bottom-right block, initially filled with zeros.

\begin{figure}[tb]
 \centering
 \includegraphics[width=\columnwidth]{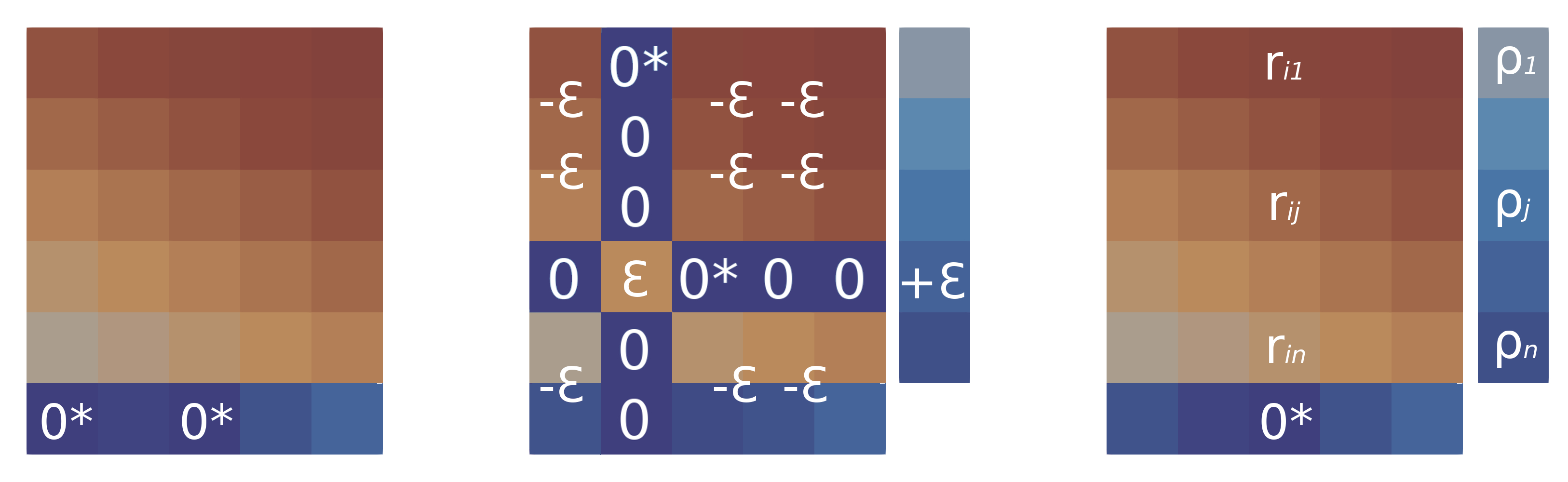}
 \caption{In our setup, every element in the last row is considered
 independent, so that it can contain multiple starred zeros (left).
 This emulates the \maxime{behavior} of the bottom-left matrix block in the
 classical approach.
 During an $\epsilon$-reduction phase (center), we keep track of the
 (always positive) quantities that were added to matrix rows,
 hence increasing the top-right block in the classical approach,
 initially filled with only \maxime{zeros}.
 If a zero is starred in the last row and $j^{th}$ column,
 let $\rho_i$ be the sum of
 quantities added to row $i$ throughout the algorithm (right).
 If for all $i$, $r_{ij} > \rho_i$, then the persistence pair
 associated with column $i$ is assigned to the diagonal. 
 If not (which never happened in our experiments),
 row residuals $\rho_i$ and the equivalent residuals for columns $\rho_j$
 are used to report the partial optimization
 onto the matrix of the exact classical approach.
 }
 \vspace{-2ex}
 \label{fig:smunk}
\end{figure}

We therefore use the criterion stated in \eqrref{eq:criterion} to
ensure that if, at any given iteration
of the algorithm, a zero is starred in the last row of
column $j$, the cost of assigning the
corresponding persistence pair
to any other pair is higher than the cost of leaving both unassigned
(0 for the $j^{th}$-column pair and the \emph{residual value}
$\rho_i$ for $i^{th}$-row pairs -- see \algref{alg:sparse}).
This specificity 
is illustrated in \figref{fig:smunk}.

\vspace{-2ex}
\begin{equation}
\forall i \in \llbracket0, n\rrbracket, r_{i, j} > \rho_{i} \Rightarrow r_{n+1,j} = 0^*
 \label{eq:criterion}
\end{equation}
\vspace{-3ex}

The \eqrref{eq:criterion} criterion is checked whenever a zero appears on the last row after
a subtraction is performed on a column by the algorithm. If it is observed, the
corresponding column is removed from the problem and the persistence pair is set unassigned.

If the criterion is not respected, we have to report back the reduced problem
onto the full matrix (missing banned columns and with reported found \emph{residuals}
$\rho$).
For this\maxime{,} we need to keep track of residuals, that is, values that have been added
or subtracted from each row and column throughout the course of the algorithm.
Once these residuals have been reported onto the full matrix, there can be no negative
element, and all of the optimization work has been reported (so that
we do not start all over again from the beginning, but we start from the \maxime{optimized}
output of the first phase).

In practice for persistence diagrams, we always observed that the first phase is
sufficient to find an optimal assignment.
\maxime{Using this approach prevents from}
working with
two potentially large blocks of persistence elements, typically occurring with the
complete matrix for $i\in \llbracket n+1, m+n\rrbracket $ and $j \in \llbracket m+1, m+n\rrbracket $.
This property is further motivated by the use of geometrical lifting
(\secref{sec:liftedPersistenceMatching}).
The approach is detailed in \algref{alg:sparse}.


\MyBox[-8ex]{startc}{endc}{blue}{6ex}
\MyBox[-3.5ex]{startb}{endb}{blue}{1ex}
\MyBox[-0.3ex]{starta}{enda}{blue}{1ex}

\begin{algorithm}[t]
 \KwData{$R'=(r_{ij})$, an  
  \blueish{$(n+1) \times m$ persistence}
  cost matrix,\\
  \blueish{$R$ the full $(n+m)^2$ matrix with non-sparse blocks.}
  }

 \KwResult{$S$ a set of starred independent zeros}
 \tikzmark{starta}$\forall i,  \rho_i \leftarrow 0$ // row residuals\\
 $\forall j, \rho_j \leftarrow 0$ // column residuals\\
 $B \leftarrow \emptyset$ // banned columns\\
 Subtract the persistence element from every row and $\rho_i$\tikzmark{enda}\\
 Transpose $R'$ if $n > m$ and let $k=\min(m, n)$\\
 Subtract the min element from every \blueish{column} 
  of $R'$ \blueish{and $\rho_j$} \\
 Star independent zeros and cover their columns\\

 \While{number of covered columns $< k$}{
  Find a non-covered zero $Z_1$ and prime it\\
  \eIf{ \blueish{$Z_1$ is in the last row or} there is no 0* in its row}{
    Augmenting path phase (\figref{fig:augment})\\
    Erase all primes, reset all covers\\
    Cover each column of containing a starred zero\\
  }{
    Let $Z_1'$ be the 0* in the row of $Z_1$\\
    Cover this row and uncover the column of $Z_1'$\\
  }
  \If{there is no uncovered zero left}
  {
    Matrix $\epsilon$-reduction phase (\figref{fig:reduce})\\
    \tikzmark{startc}
    $\rho_i \leftarrow \rho_i + \epsilon$ for modified rows $i$ \\
    $\rho_j \leftarrow \rho_j - \epsilon$ for modified columns $j$ \\
      \If{$\exists j | r_{n+1,j} = 0$ and
          $\forall i \in \llbracket1, n\rrbracket, r_{i, j} > \rho_i$\tikzmark{endc}}{
          $r_{n+1,j}$ is starred\\
          $B \leftarrow B \cup j$\\
      }
  }
 }
  \tikzmark{startb}\If{$\exists j \notin B | r_{n+1,j} = 0^*$ and $\exists i| r_{i, j} < \rho_i$}{
      Kuhn-Munkres($R_{ij}'' = R_{ij}+\rho_i+\rho_j$) with $j \notin B$.\tikzmark{endb}
  }
 \caption{Our algorithm for sparse persistence matching.
 Blue sections 
 allow to emulate the behavior of the three original
 non-sparse blocks on one single row,
 while ensuring optimality thanks to the residuals column.
 Black sections are common with the unbalanced
 Kunk-Munkres algorithm.
 \vspace{-2ex}
 }
 \label{alg:sparse}
\end{algorithm}

\subsection{Sparse assignment}
\label{sec:sparseAssignment}

\maxime{In practice, it is often observed that}
some assignments are not possible,
and that reordering columns in the associated cost matrix
would enable faster lookups and modifications \cite{cui16}, using sparse
matrices.
With persistence diagrams, the following
simple criterion \maxime{(\eqrref{eq:cancellation})} can be used to discard
lookups for potential matchings.

\vspace{-1ex}
\begin{equation}
 d_{\nu}(p, q) > d_{\nu}(p, \maxime{\text{diag}}(p)) +
 d_{\nu}(q, \maxime{\text{diag}}(q))
 \label{eq:cancellation}
\end{equation}
\vspace{-3ex}

\begin{figure}[ht]
 \centering
 \includegraphics[width=\columnwidth]{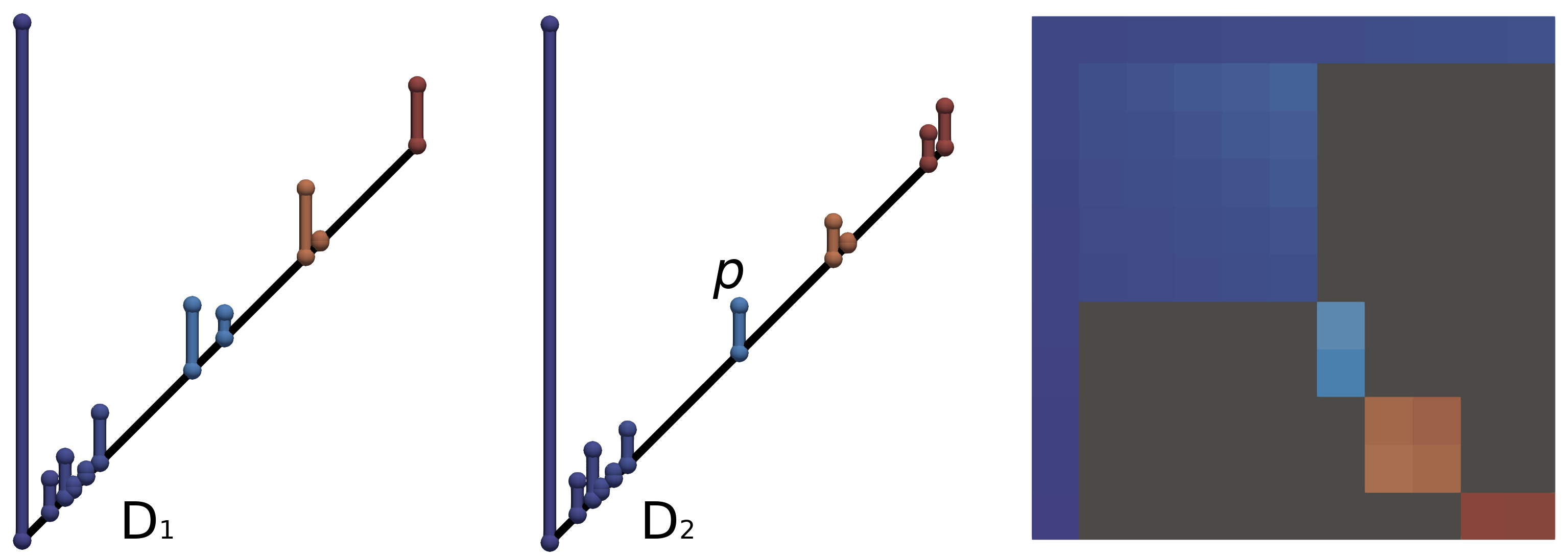}
 \caption{Persistence diagrams $D_1$ and $D_2$ showing in color
 small persistence pairs that
 will never be assigned in an optimal matching. The light blue pair $p \in D_2$ is such that
 $d(p, \maxime{\text{diag}}(p)) + d(q, \maxime{\text{diag}}(q)) < d(p, q)$ for any $q \in D_1$ which is neither
 light blue nor the first large persistence pair.
 This results in the cost matrix (right, $D_1$ pairs are rows and $D_2$
 pairs are columns), where \maxime{gray} elements correspond to pairs $(p,q)$ s.t.
 $d(p,q) > d(p,\maxime{\text{diag}}(p))+d(q,\maxime{\text{diag}}(q))$.
 }
 \vspace{-2ex}
 \label{fig:sparse}
\end{figure}

Working with our version of the Kuhn-Munkres algorithm then becomes interesting
for many assignments \maxime{verify \eqrref{eq:cancellation}}
(\figref{fig:sparse}),
hence greatly reducing the lookup time
for zeros, minimal elements, and the access time for operations
performed on rows or columns.

\maxime{On the contrary, t}he original full-matrix version of Kuhn-Munkres
deals with non-sparse
blocks which have to be accessed and modified constantly
throughout the course of the algorithm.

\begin{figure}[tb]
 \centering
 \includegraphics[width=\columnwidth]{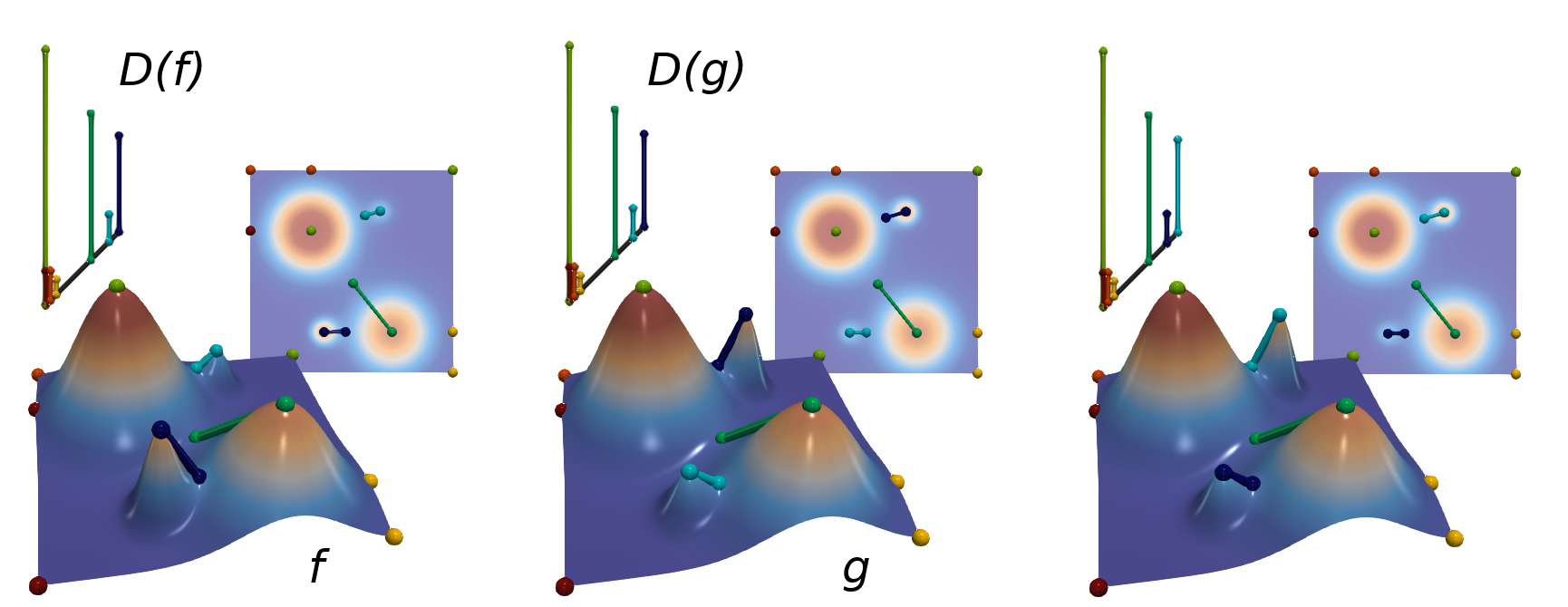}
 \caption{Scalar field $f$ with persistence diagram $D(f)$ (left),
 matched with a scalar field $g$ with a similar persistence diagram $D(g)$,
 but an embedding that swaps the position of
 the light blue pair with that of the dark blue pair.
 Matched pairs are displayed with the same \maxime{color} using the
 non-geometric (center) and geometric Wasserstein metric (right).
 The latter takes the geometrical embedding into account,
 preventing similar pairs (regarding persistence) to be assigned
 if they are geometrically distant.}
 \label{fig:mismatch}
 \vspace{-2ex}
\end{figure}

\section{Lifted persistence Wasserstein metric}
\label{sec:liftedPersistenceMatching}
This section highlights the limitations of the natural Wasserstein metric
applied to time-varying persistence diagrams and presents an
extension that enhances its geometrical stability.
Geometrical considerations are motivated\maxime{,} in terms
of accuracy and performance\maxime{.}

Persistence diagrams can be embedded into the geometrical domain
(\figref{fig:persistence2D}). Doing so, one easily sees
how different embeddings can correspond to similar
persistence diagrams in the \emph{birth}-\emph{death} space.
Working in this 2D space does yield irrelevant matchings:
as can be seen in \figref{fig:mismatch}, when only
the \emph{birth}-\emph{death} coordinates of persistence pairs
are considered, a matching can be optimal even if it happens
between geometrically distant zones.
As a consequence, the distance metric between
persistence pairs must be augmented with geometrical considerations.



To address this, 
we propose instead of $d_{\nu}$ (\eqrref{eq:dnu}) to use
the distance defined in \eqrref{eq:lifted}:

\vspace{-3ex}
\begin{equation}
d_{lift, \nu}(p,q) =
  (\alpha\delta_{birth}^{\nu} + \beta\delta_{death}^{\nu} +
  \gamma_1\delta_x^{\nu} + \gamma_2\delta_y^{\nu} + \gamma_3\delta_z^{\nu})
  ^{1/\nu} \\
\label{eq:lifted}
\end{equation}
\vspace{-2.5ex}

\noindent
where $\delta_x$, $\delta_y$ and $\delta_z$ correspond to geometric distances
between
the extrema involved in the
persistence pairs on a given axis.
We process diagonal projections as follows (\eqrref{eq:liftedDiagonal}):

\vspace{-2.5ex}
\begin{equation} 
\begin{split} 
d_{lift, \nu}(p, \maxime{\text{diag}}(p)) =
  & (\alpha \left|p_x\right|^{\nu} + \beta \left|p_y\right|^{\nu} + \\
  &\gamma_1(\delta_x^p)^{\nu} + \gamma_2(\delta_y^p)^{\nu} +
  \gamma_3(\delta_z^p)^{\nu})^{1/\nu}
\label{eq:liftedDiagonal}
\end{split}
\end{equation}
\vspace{-2ex}

\noindent
where the terms $\delta_x^p$, $\delta_y^p$ and $\delta_z^p$ correspond to the
geometric distance between the critical points of
a given pair $p$.
Intuitively, it accounts for the distance between the critical points to cancel.

\begin{figure}[tb]
 \centering
 \includegraphics[width=\columnwidth]{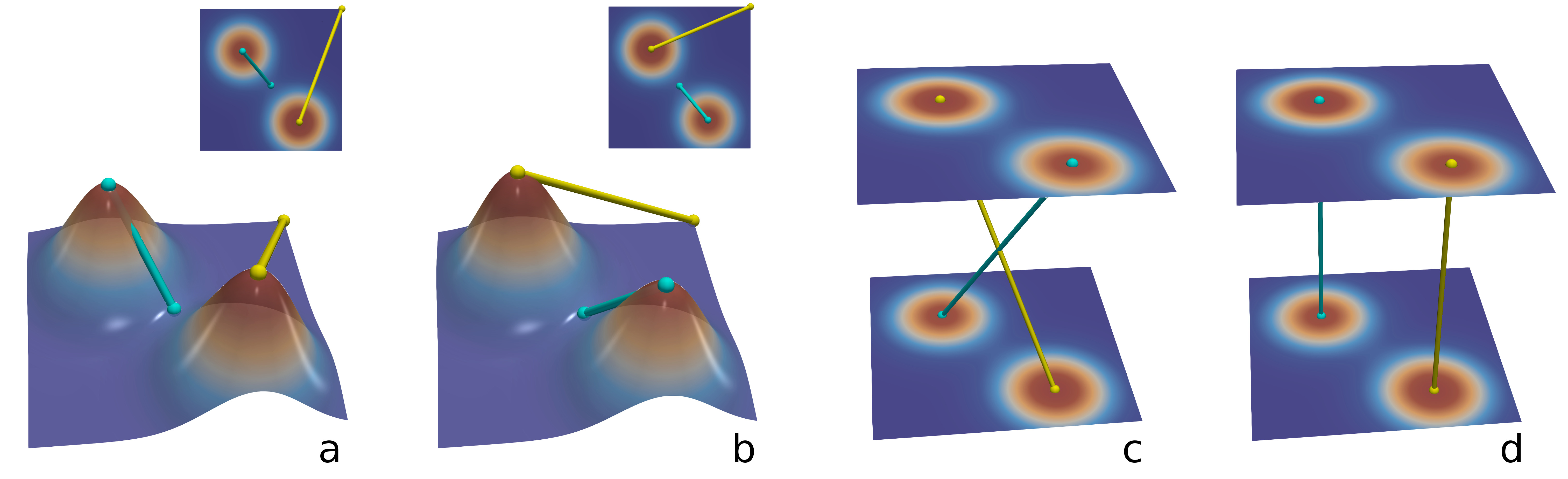}
 \caption{Lifting the birth coordinate.
 2D scalar fields with two gaussians (a, b), where the bottom (resp. top) gaussian
 has the maximum value (a) (resp. b). Using the geometrical metric alone (c)
 is not sufficient, as the \emph{birth} coordinate
 $p_x$ misleadingly equalizes the persistence term of pairs of the same \maxime{color} in (a, b):
 $\delta_{p, yellow}^a = \delta_{p, yellow}^b$, $\delta_{p, blue}^a = \delta_{p, blue}^b$,
 potentially overcoming the geometrical factor.
 Lifting the birth coordinate with a small coefficient for associating maxima yields
 the correct matching (d).
 }
 \vspace{-2ex}
 \label{fig:lift-birth}
\end{figure}

A \emph{lifted} distance is considered by augmenting the geometric
distance with coefficients $\alpha, \beta, \gamma_i$.
This aims \maxime{at} giving more or less importance
to the birth, death or some of the $x,y,z$ coordinates during the matching,
depending on applicative contexts.
For instance, \maxime{in practice} it is desirable to give less importance to
the birth coordinate when dealing with $d$-$(d-1)$ persistence pairs (in other
words, for tracking local maxima, see \figref{fig:lift-birth}).
For the remainder of the paper and the experiments, we used
$(\alpha, \beta, \gamma_i) = (0.1, 1, 1)$ for maxima and
$(\alpha, \beta, \gamma_i) = (1, 0.1, 1)$ for minima,
for normalized geometrical extent and scalar values.
We observed that using a lifted metric further
increases the \maxime{cost matrix} sparsity, resulting in extra speedups.
%

\begin{figure*}[tb]
 \centering
 \includegraphics[width=\linewidth]{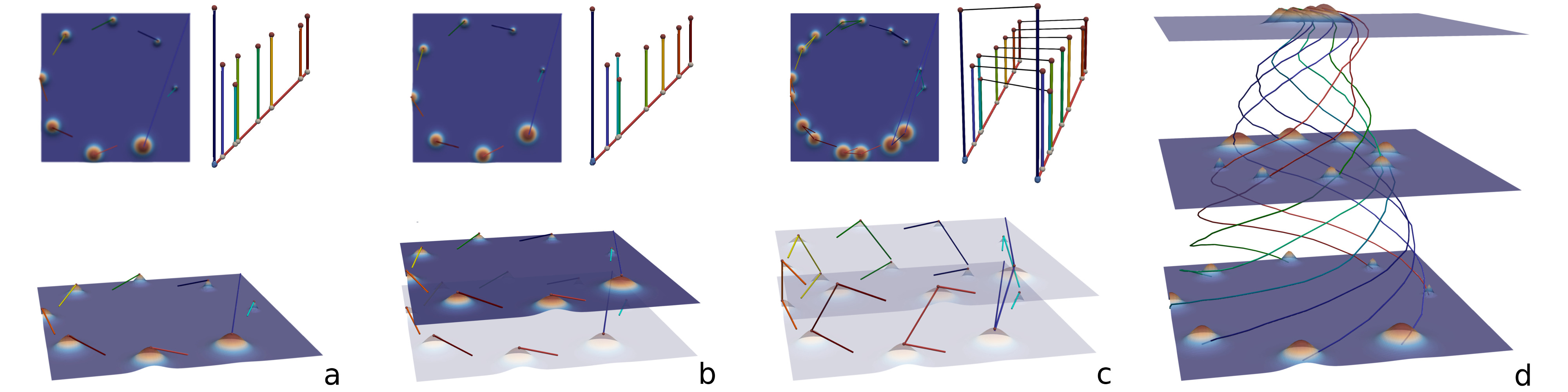}
 \caption{Overview of our tracking approach
  on a dataset consisting of eight whirling gaussians:
  persistence diagram computations for two consecutive timesteps (a) and (b);
  matching of persistence pairs of two timesteps (c),
  propagation of matchings and construction of a trajectory (d).}
 \label{fig:overview}
\end{figure*}


\section{Feature tracking}
\label{sec:featureTracking}

This section describes the four main stages of our tracking framework,
relying on the discussed theoretical setup.
Without loss of generality, we assume that the input data is
a time-varying 2D or 3D scalar field defined on a PL-manifold.
Topological features are extracted for all timesteps
(\figref{fig:overview}, a-b), then matched (\figref{fig:overview}, c);
trajectories are built from the successive matchings
(\figref{fig:overview}, d) and post-processed to detect merging and splitting events.

\subsection{Feature detection}
\label{sec:featureMatching}

First, we compute persistence diagrams for each timestep.
\maxime{We propose} using the algorithm by Gueunet et al. \cite{gueunet_ldav17}\maxime{,
in which} only $0$-$1$ and $d$-$(d-1)$ persistence pairs \maxime{are considered}.

When the data is noisy, it is possible to discard pairs of low persistence
(typically induced by noise) by applying a simple threshold.
In practice, this amounts to 
only considering the most prominent features.
Using such a threshold 
\maxime{accelerates} the matching process, where for approaches based on overlaps,
removing topological noise would require a topological simplification of
the domain (for example using the approach in \cite{tierny_vis12}),
which is computationally expensive.

\subsection{Feature matching}

If $P_1, P_2$ are two sets of persistence pairs taken at timesteps
$t$ and $t+1$, then we use the algorithm described in \secref{sec:persistenceCriterion},
with the appropriate distance metric, as discussed in \secref{sec:liftedPersistenceMatching},
to associate pairs in $P_1$ and $P_2$.
A given pair $p_1 \in P_1$ might be associated to one pair $p_2 \in P_2$ at most,
or not associated, and \maxime{symmetrically}.

\subsection{Trajectory extraction}

Trajectories are constructed by simply attaching successively matched segments.
For all timesteps $t$, if the feature matching associates
$p_i$ with $M_t(p_i) = q_j$, then a segment is traced between $q_j$ and $p_i$,
and is potentially connected back to the previous segment of $p_i$'s trajectory.
If $M_t(p_i) = \maxime{\text{diag}}(p_i)$, the current trajectory ends.
\maxime{Properties are associated to trajectories (time span, critical point index)
and to trajectory segments (matching cost, scalar value, persistence value,
embedded volume).}

\subsection{Handling merging and splitting events}
\label{sec:mergeSplit}

\begin{figure}[tb]
 \centering
 \includegraphics[width=\columnwidth]{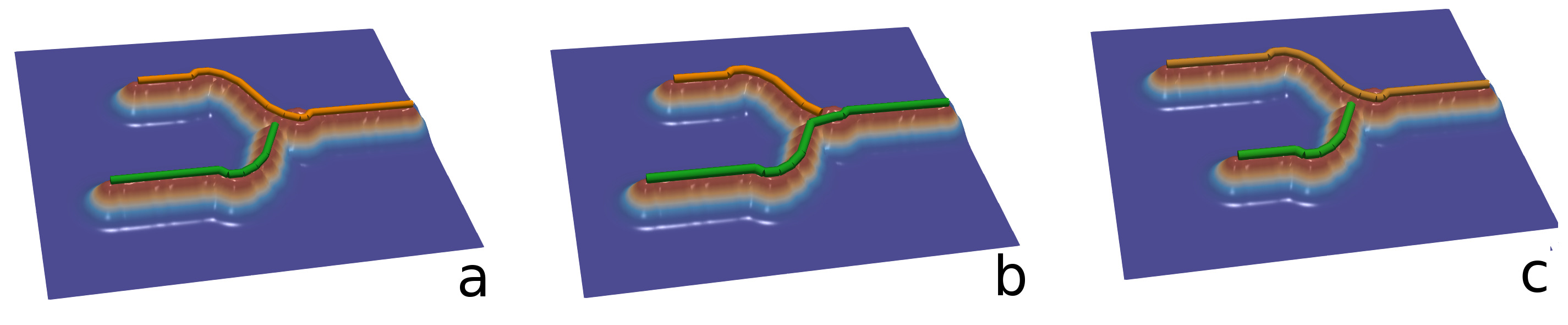}
 \caption{Merging process. Tracking is performed on two
 gaussians moving from left to right (a). 
 The post-process
 detects a merging event with a geometrical threshold, then
 propagates the component identifier of the oldest component (b)
 and properly reconnects the matching segment.
 If the oldest component has already the right identifier (c)
 nothing is done.
 This process is proposed by analogy with building
 persistence diagrams.
 }
 \vspace{-2ex}
 \label{fig:merge}
\end{figure}

Given a user-defined geometrical threshold $\epsilon$,
we propose to detect events of \emph{merging} or \emph{splitting}
along trajectories in the following manner.
If $T_1, T_2: I\subset \mathbb{N} \rightarrow \mathbb{R}^3 $ are two trajectories spanning
throughout $[t_i, t_{i+n}]$ and $[t_j, t_{j+m}]$ respectively,
and if for some $k \in [i, i+n] \cap [j, j+m]$, $d_{lift, \nu}(T_1(t_k), T_2(t_k)) < \epsilon$,
where \maxime{$d_{lift, \nu}$ is a lifted} distance,
then an event of merging (or splitting) is detected.
We consider that a merging event occurs between $T_1$ and $T_2$ at time $k$,
when neither $T_1$ nor $T_2$ start at $t_k$. We then consider that
the \emph{oldest} trajectory takes over the \emph{youngest}.
For example, $T_1$ and $T_2$ meet (according to the $\epsilon$ criterion)
at $t_k$ the last timestep of $T_2$, and $T_2$ started \emph{before} $T_1$, then
we disconnect the remainder of $T_1$ from the trajectory before $t_k$ and
we connect it so as to continue $T_2$ until $T_1$'s original end.
Similarly, a splitting event occurs between $T_1$ and $T_2$ at time $k$,
when neither $T_1$ nor $T_2$ end at $t_k$.
The process is illustrated in \figref{fig:merge}. \maxime{It is done 
separately for distinct critical point types: minima, maxima and saddles
are not mixed.}

\section{Results}
\label{sec:results}

This section presents experimental results obtained on a desktop
computer with two Xeon CPUs (3.0 GHz, 4 cores each), with 64 GB of RAM.
We report experiments on 2D and 3D time-varying datasets, that were either
simulated with Gerris \cite{Popinet03}
(von K\'arm\'an Vortex street, Boussinesq flow,
starting vortex), or acquired (Sea surface height, Isabel hurricane).
Persistence diagrams are computed with the implementation of
\cite{gueunet_ldav17}
available in the Topology ToolKit \cite{ttk};
the tracking is restricted to $0$-$1$ and $(d-1)$-$d$ pairs.
We implemented our matching (\secref{sec:persistenceCriterion}) and tracking 
approaches (\secref{sec:featureTracking}) 
in C++ as a Topology ToolKit module. 



\subsection{Application to simulated and acquired datasets}
\label{sec:application}

\begin{figure}[tb]
 \centering
 \includegraphics[width=\linewidth]{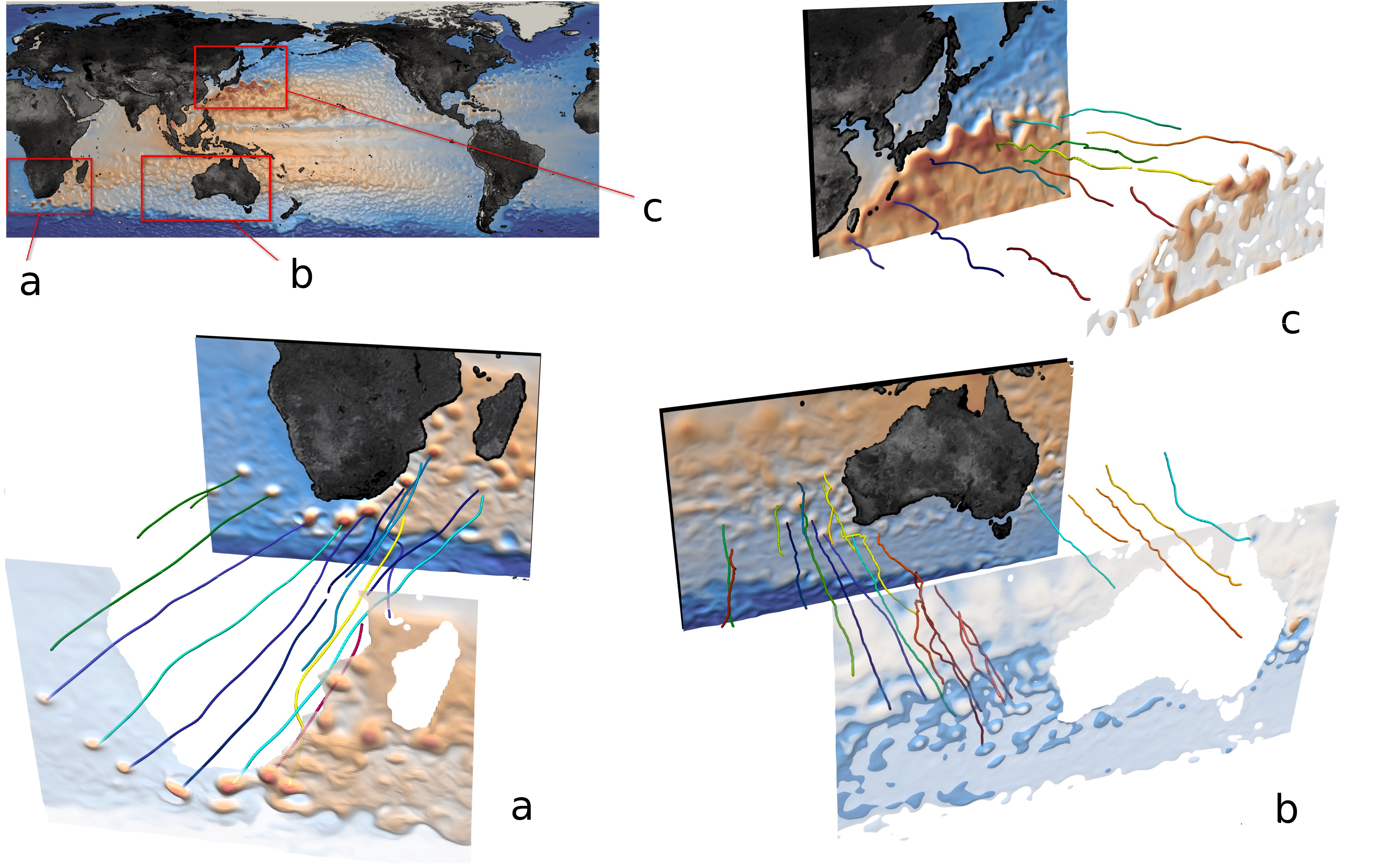}
 \caption{Sea surface height (SSH) captured over 365 days,
 1 timestep every day. Local maxima are tracked in the region
 corresponding to the Agulhas Current, near South Africa (a);
 it is observed they are slowly drifting towards the west.
 SSH maxima are also drifting west
 in the less contrasted zone of the West Australian Current (b).
 Tracking in the region of the Kuroshio Current,
 near Japan (c), demonstrates a whirling \maxime{behavior} of local
 maxima.
 }
 \vspace{-2ex}
 \label{fig:elevation}
\end{figure}

\begin{figure}[tb]
 \centering
 \includegraphics[width=0.9\columnwidth]{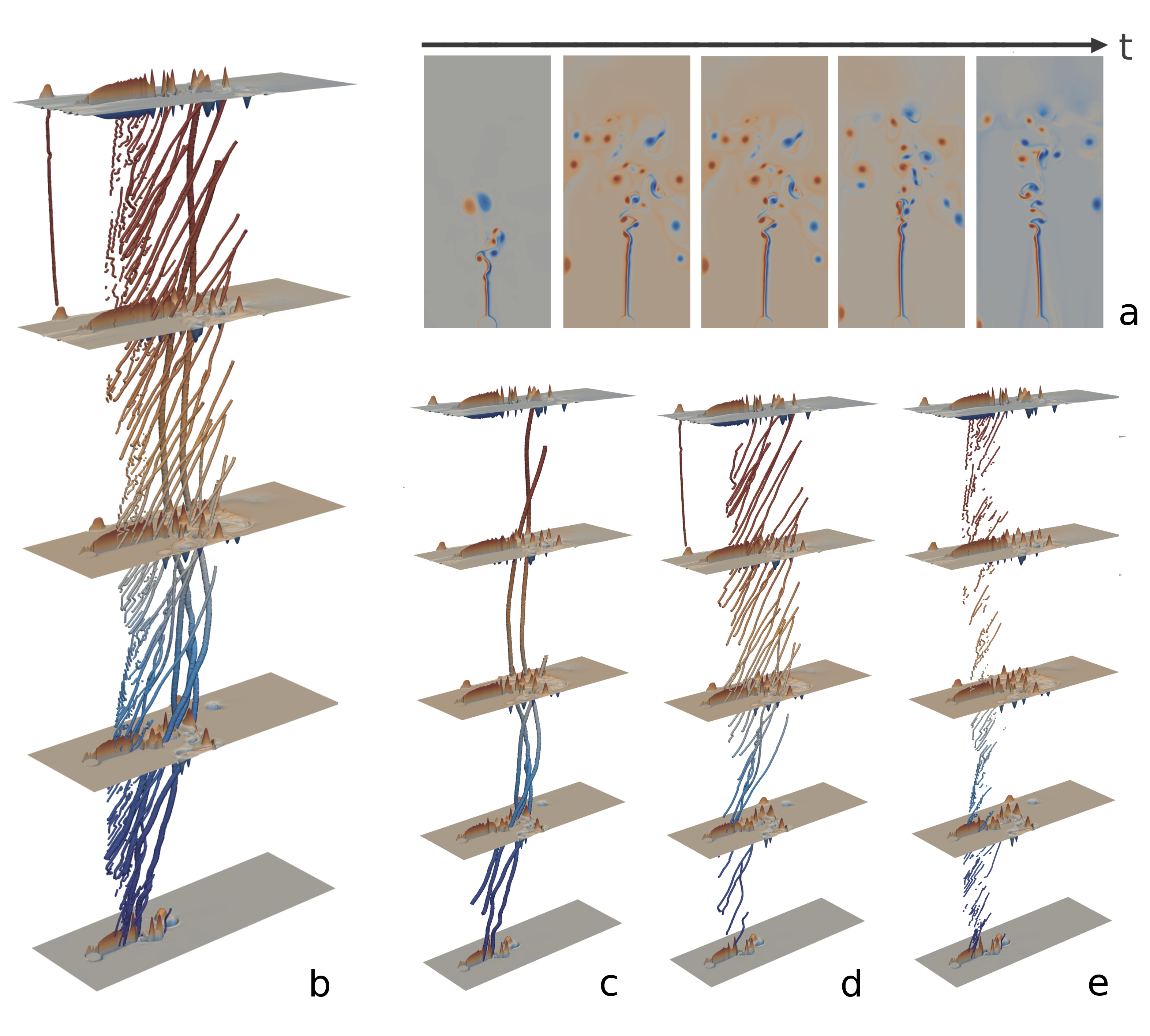}
 \caption{Boussinesq flow generated by a heated  cylinder (a).
 Feature tracking is performed (b) on the fluid vorticity.
 Some vortices exist over a long period of time (c),
 as others vanish more rapidly (d), sometimes akin to noise (e).
 Feature trajectories can easily be filtered from their lifespan.
 }
 \vspace{-2ex}
 \label{fig:boussinesq}
\end{figure}

\begin{figure}[tb]
 \centering
 \includegraphics[width=\columnwidth]{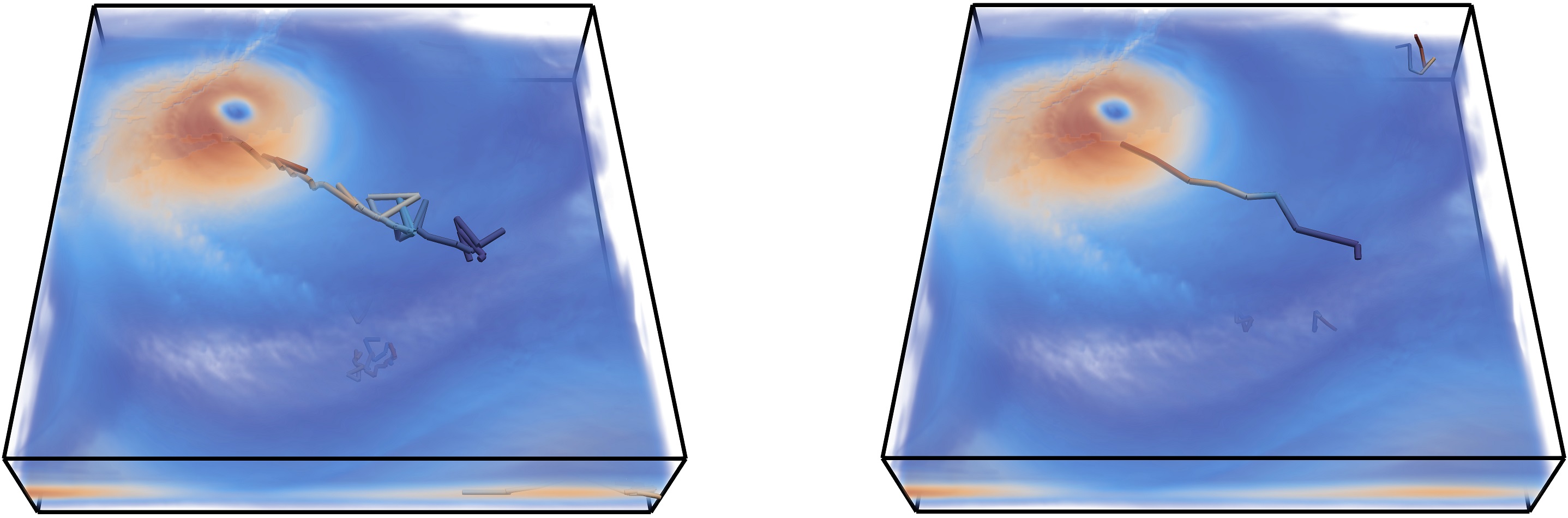}
 \caption{Tracking performed on the wind \maxime{velocity} on a
 3D Isabel hurricane dataset
 before (left), and after (right) temporal downsampling
 (1 frame every 5 timesteps). The global maximum is
 tracked successfully despite the high
 \maxime{instability} displayed by the scalar field.}
 \vspace{-2ex}
 \label{fig:isabela}
\end{figure}

\begin{figure}[tb]
 \centering
 \includegraphics[width=\columnwidth]{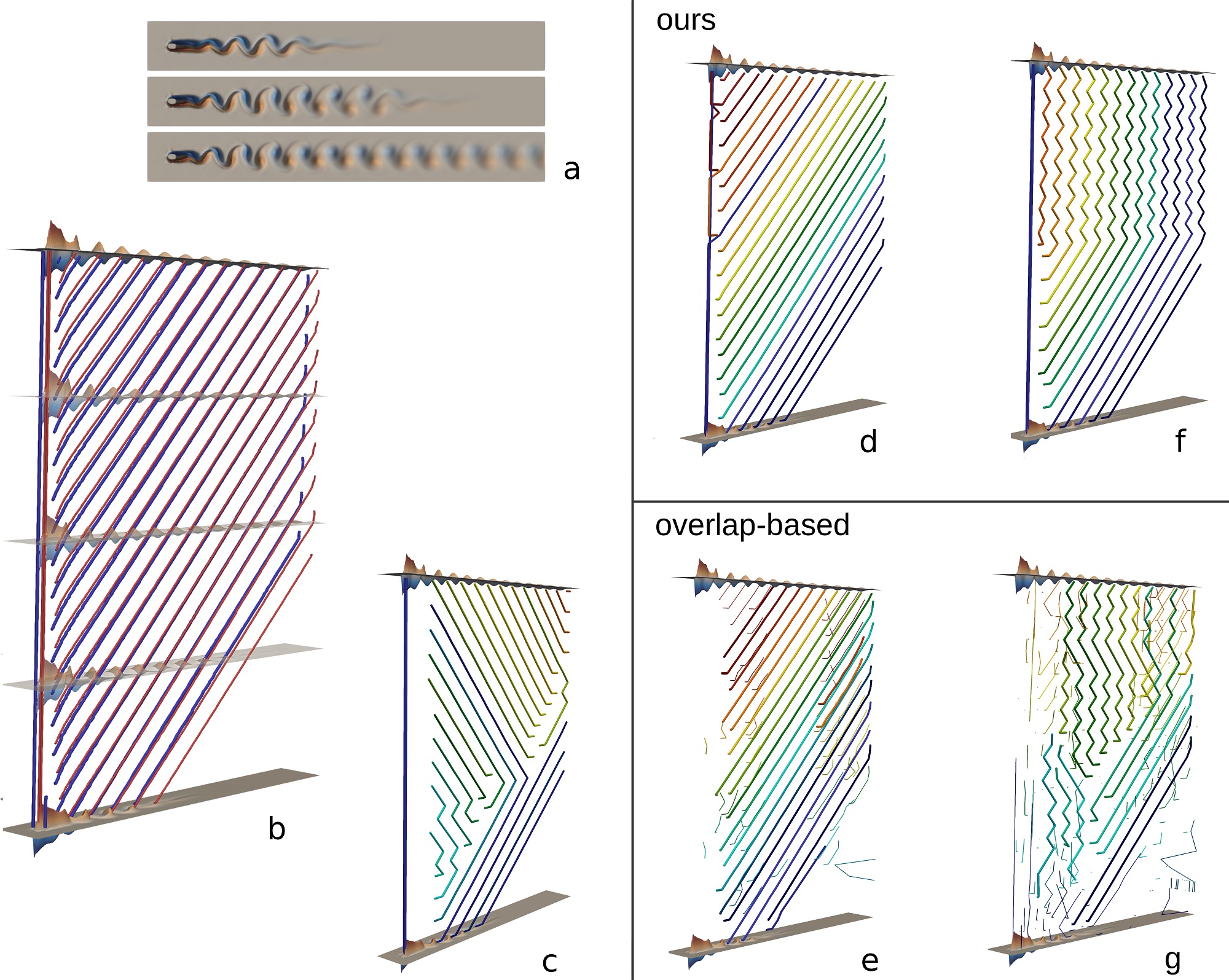}
 \caption{Simulated von K\'arm\'an vortex street (a),
 on which minima and maxima of the vorticity
 are tracked with our approach
 and 1\% \maxime{persistence} filtering (b).
 Only taking the geometry and scalar value into account
 while doing the matchings
 (i.e. completely ignoring the \emph{birth} in the lifted metric),
 is not sufficient to correctly track features (c).
 Maxima only are tracked
 considering 1 frame every 5 timesteps (d). With the
 same temporal resolution,
 the overlap-based approach (e) does capture small trajectories
 corresponding to noise, displayed with thinner lines,
 that have to be filtered for instance
 using topological simplification \cite{tierny_vis12}.
 Considering 1 frame every 7 timesteps (f) still
 yields correct trajectories up to the point where, every
 other frame, optimal matchings for the metric
 are between a feature and the preceding one,
 due to features \maxime{traveling} fast.
 The overlap approach (g) is less stable in this case
 as it extracts erroneous trajectories from the very first
 stages of the simulation to the end.
 }
 \vspace{-2ex}
 \label{fig:vortexstreet}
\end{figure}

We applied our tracking framework to both
simulated and acquired time-varying datasets
to outline specific phenomena.

In \figref{fig:elevation}, we present the results
of the tracking framework applied to an oceanographic
dataset. The scalar field (sea surface height) is defined
on 365 timesteps on a triangular mesh. We can see
interesting trajectories corresponding to well-known oceanic currents.
Drifting (a, b) and turbulent \maxime{behaviors} (c) of local extrema are highlighted.
In \figref{fig:boussinesq}, tracking is performed
on the vorticity of highly unstable Boussinesq flow.
Thanks to our analysis, trajectories can be filtered according
to their temporal lifespan, revealing clearly different trajectory
patterns among the turbulent features.
This kind of \maxime{analysis} may be easily performed based on other
trajectory attributes, depending on applicative contexts.
In \figref{fig:isabela}, we show our approach on a 3D hurricane 
dataset whose temporal resolution is such that a method based
on overlaps of split-tree leaves (see \secref{sec:trackPerf})
could not extract trajectories.
In \figref{fig:vortexstreet}, our tracking framework correctly follows
local extrema of the vorticity field in a simulated vortex street.


\subsection{Tracking robustness}

In the following two sections, we demonstrate the robustness
and performance of our tracking framework.

We compare to the greedy approach based on the overlap of volumes
\cite{bremer10, bremer_tvcg11, bajaj06, Saikia17} of
split-tree leaves, which amounts to tracking local maxima.
In this approach, for every pair of consecutive timesteps $(t,t+1)$,
split-tree segmentations $S_t$ and $S_{t+1}$ are computed
(these are a set of connected regions).
Overlap scores are then computed
for every pair of regions $(s_i,s_j) \in S_t \times S_{t+1}$,
as the number of common
vertices between $s_i$ and $s_j$.
Scores are sorted and $s_i$ is considered matched
to the highest positive scoring $s_j$
such that $s_j$ has not been matched before.
Trajectories are \maxime{extracted} by repeating the process for all timesteps.

\begin{figure}[tb]
 \vspace{-2ex}
 \centering
 \includegraphics[width=\columnwidth]{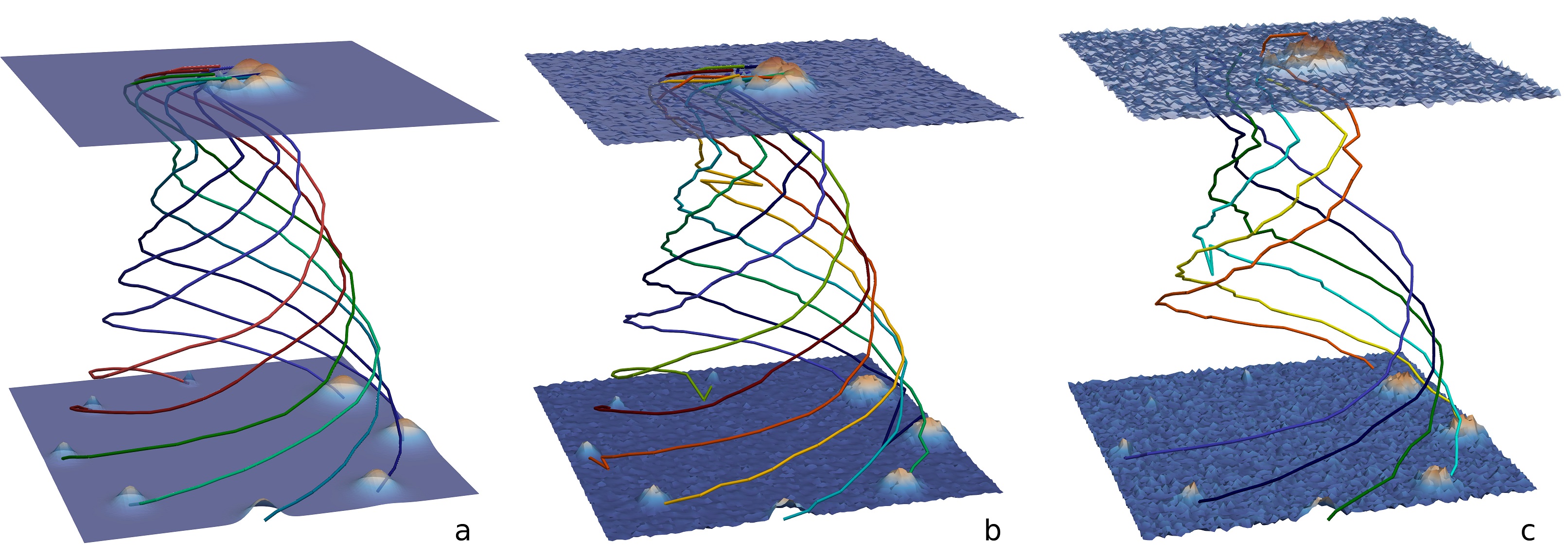}
 \caption{Lifted Wasserstein tracking performed on
 a set of whirling 2D gaussians (a).
 With noise accounting for 10\% of the scalar range (b),
 feature trajectories are still correctly detected.
 For 25\% noise (c), 75\% of the features (namely, the 6
 most prominent out of the initial 8)
 are still correctly tracked despite heavy perturbations.
 }
 \vspace{-2ex}
 \label{fig:noise}
\end{figure}

The robustness of our tracking framework
is first assessed on a synthetic dataset consisting of
whirling gaussians, on which we applied noise (\figref{fig:noise}).
Identified trajectories are sensibly the same with a perturbation
of 10\% of the scalar range.
The 75\% most important features are still correctly tracked
after a 25\% random perturbation has been applied to the data.

In \figref{fig:vortexstreet},
our method is compared to the
greedy approach, based on overlaps, while decreasing the
temporal resolution.
The overlap approach yields trajectories
corresponding to noise (\figref{fig:vortexstreet}-e),
which can be filtered by applying topological simplification
beforehand (this would have a significant computational cost
as it requires to modify the original function),
or by associating the scalar value of the function to
every point in the trajectory and then filtering the trajectory
in a post-process step.
In our setup, is is much simpler to discard this noise,
by using a threshold for discarding small
persistence pairs before the matching (implying a faster matching
computation).
When downsampling the temporal resolution to only 20\% of the timesteps,
our approach still gives the correct results (\figref{fig:vortexstreet}, d vs. e).
With 15\% of the timesteps, our approach (\figref{fig:vortexstreet}-f)
still agrees with the tracking on the
full-resolution data (\figref{fig:vortexstreet}-b),
until preceding features begin to catch up, resulting
in a zig-zag pattern.
By comparison, the overlap method fails to correctly track
meaningful regions from the beginning of the simulation to its end;
it is indeed dependent on the
geometry of overlaps, which is unstable.
It can be argued that the locality captured by overlaps
is emulated in our framework by 
embedding and lifting 
the Wasserstein metric, when
the overlap method does not take
persistence into account when matching regions.
Also note that if the saddle component of persistence pairs
associated to maxima is ignored 
(i.e. if $\alpha=0$ in \eqrref{eq:lifted} and \eqrref{eq:liftedDiagonal})
during the matching, then the geometrical distance can be 
insufficient for correctly tracking these persistence pairs (c).
Therefore, the problem of matching persistence pairs for tracking topological
features cannot be reduced to the (unbalanced) problem of
assigning critical points in 4 dimensions (3 for the geometrical extent,
one for the scalar value).

\figref{fig:isabela} further illustrates the robustness of
our approach when downsampling the data temporal resolution.
\maxime{In h}urricane datasets\maxime{,}
local maxima can be displaced to geometrically distant zones
between timesteps if those are taken at multiple-day intervals.
This unstable \maxime{behavior} and the absence of obvious overlaps
makes it particularly difficult to track extrema;
nevertheless, our framework managed to track
\maxime{them} at a very low time-resolution.


\subsection{Tracking performance}
\label{sec:trackPerf}

We then compare our framework with our implementation of the 
approach based on overlaps
\cite{bremer_tvcg11}
on the ground of performance.
Figures are given in \tabref{tab:greedy-comparison}.
Note that our approach has the advantage of taking persistence diagrams
as inputs, so it can be applied to unstructured or time-varying meshes,
for which \maxime{overlap computations are} not trivial.
Our approach is also relatively dimension-independent: though in 3D,
computing overlaps is very time-consuming
(\figref{tab:greedy-comparison}-Isabel),
the complexity of the Wasserstein matcher,
which only takes embedded persistence diagrams as inputs,
for a given number of persistence pairs
is sensibly equivalent.
For both Isabel and Sea surface height datasets,
we applied a 4\% persistence filtering on input persistence diagrams.
As the experiments show, our approach is faster in practice than the
overlap method \maxime{with best-match search}.

\begin{table}[hb]
    \centering
    \caption{
    Time performance comparison (CPU time in seconds)
    between the approach based on
    overlaps of volumes \cite{bremer_tvcg11} and our lifted Wasserstein
    approach.
    Tracking is performed over 50 timesteps,
    on \maxime{structured} 2D (Boussinesq, Vortex street),
    structured 3D (Isabel), and unstructured 2D (Sea surface height)
    meshes.
    The pre-processing step (FTM)
    computes
    the topology of the dataset. 
    The post-processing
    step extracts the tracking mesh, computes its attributes,
    and handles splitting and merging events.
    We observe a parallel speedup ranging from 4 to 6
    for our approach on 8 threads (FTM and matching phases).
    }
    \label{tab:greedy-comparison}
    \scalebox{0.85}{
    \centering
    \begin{tabular}{|l|r|rr|r|}
    \toprule
    Data-set
    & \multicolumn{1}{c}{Pre-proc (s)}
    & \multicolumn{2}{|c|}{Matching (s)}
    & \multicolumn{1}{c|}{Post-proc} \\
    & \multicolumn{1}{c}{FTM}
    & \multicolumn{1}{|c}{\cite{bremer_tvcg11}} & ours 
    & \multicolumn{1}{c|}{(s)} \\
    \midrule
    Boussinesq    & 116 & 75    & 18   
                                              & 4.7 \\
    Vortex street & 45  & 23    & 18   
                                              & 2.8 \\
    Isabel (3D)   & 863 & $>$3k & 17   
                                              & 162 \\
    Sea height    & 568 & N.A.  & 277  
                                              & 113 \\
    \bottomrule
    \end{tabular}
    }
\end{table}


\subsection{Matching performance}
\label{sec:matchingResults}

Next, we compare the performance of the matching method we
introduced in \secref{sec:persistenceCriterion}
to two other state-of-the art algorithms.

\begin{table}[tb]
    \centering
    \vspace{-2ex}
    \caption{
    Time performance comparison between the state-of-the-art Munkres-based
    approach \cite{weaver13}, 
    and our modified sparse approach.
    }
    \label{tab:comparison}
    \label{tab:timeComparison}
    \scalebox{0.85}{
    \centering
    \begin{tabular}{|ll|rr|}
    \toprule
    Data-set
    & Sizes of diagrams & \multicolumn{2}{c|}{Time (s)} \\
    && \cite{weaver13} 
     & ours
     \\
    \midrule
    Starting vortex    & $473 - 489$   & 68.6
                                          & 1.26
                                          \\
    Isabel             & $465 - 413$   & 72.2
                                          & 3.58
                                          \\
    Boussinesq         & $1808 - 1812$ & 11.1k
                                          & 102
                                          \\
    Sea height         & $1950 - 5884$ & 26.5k
                                          & 155
                                          \\
    \bottomrule
    \end{tabular}
    }
\end{table}

\begin{figure}[tb]
 \centering
 \includegraphics[width=\columnwidth]{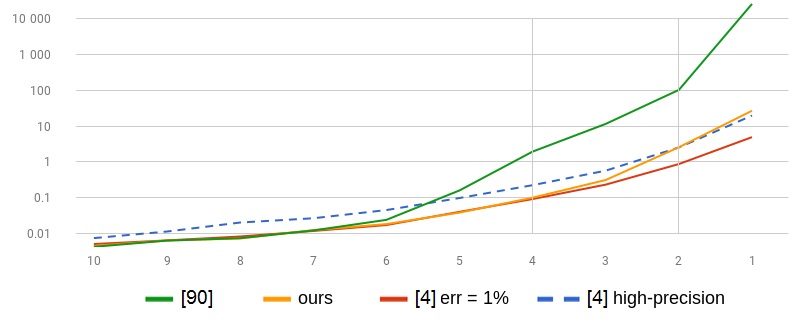}
 \vspace{-2ex}
 \caption{Running times in seconds of different
 matching approaches, for decreasing persistence thresholds
 expressed in percentage of the scalar range.
 The
 initial two diagrams 
 containing 14,082 and 14,029 pairs
 are filtered to remove pairs
 whose persistence is less than the defined threshold,
 then a matching is computed with our new method, 
 the reference exact method \cite{weaver13}, 
 the approximate method \cite{bertsekas89}, first with
 1\% accuracy, then with an accuracy of $10^{-4}$\% of the scalar range.
 }
 \vspace{-3ex}
 \label{fig:auction}
\end{figure}

We compare it to the \emph{reference approach} for the exact assignment
problem \cite{weaver13} based on the Kuhn-Munkres algorithm,
and to our implementation of the \emph{approximate approach} based on
the auction algorithm \cite{bertsekas89, kerber17}, on the ground of performance.

\tabref{tab:timeComparison} shows that our new assignment algorithm is up to two orders 
of magnitude faster than the classical exact approach \cite{weaver13}.
In particular, the best speedups occur for the larger datasets which
indicates that our approach also benefits from an improved scaling.

It is often useful in practice to discard low-persistence pairs prior to any 
topological data analysis as these correspond to noise. 
\figref{fig:auction} compares the running times of our approach, \cite{weaver13}
and \cite{bertsekas89} as more and more low-persistence pairs are
taken into account. When removing pairs whose persistence
is below 5\% of the scalar range, which is commonly accepted as a conservative 
threshold, our approach is faster than all competing alternatives. When 
considering more low-persistence features, below 4\%, our approach is 
competitive with the approximated auction approach with 1\% error. Below 2\%, 
only noise is typically added in the process. The performance of our algorithm 
becomes comparable to that of the high-precision auction approximation
although our approach guarantees exact results.






\subsection{Limitations}
\label{sec:limitations}

As we described, our framework enables the tracking of 
0-1 and $d$-$(d-1)$ persistence pairs.
It would be interesting to extend it to support the tracking of
saddle-saddle pairs (in 3D) and see its application to
meaningful use cases.

Besides, the \emph{lifting} coefficients proposed in our metric
(\eqrref{eq:lifted}) might be seen as supplementary parameters that have to be
tuned according to the dataset and applicative domain.
Nonetheless, we observed in our experiments that these parameters do 
not require fine-tuning to produce meaningful tracking trajectories.
\maxime{The extent to which these can be enhanced by fine-tuning
is left to future work.}

The lifted distance can be generalized to take other parameters, such as the
geometrical volume, mass, feature speed, into account,
and be fine-tuned to answer the specificity of various scientific domains.
Merging and splitting might also be enhanced, or given more flexibility,
for instance with additional criteria.
We also believe that the performance of
the post-processing phase can be improved.

Additionally, we believe that the approximate auction algorithm can also take
the lifted persistence
metric into account by performing Wasserstein matchings between
persistence pairs in 5 dimensions, and possibly benefit from geometry-based
lookup accelerations, as suggested in lower dimension in \cite{kerber17}.
It remains to be clarified how the quality of the matchings is
affected in practice by using an approximate matching method, and how sparsity
can enhance the research phase for the auction algorithm.

\maxime{We note that the theoretical complexity of our matching
method is, as the Munkres method, cubic; however, the 
two orders of magnitude speedups
demonstrated in our experiments allow to study more challenging datasets.
For very large case studies, the use of persistence thresholds could prove
quite helpful for controlling the computing time of matchings.
Among other non-trivial tracking methods, 
some graph matching methods are based on \emph{graph-edit distances}
\cite{gao2010survey, beketayev2014measuring}.
Their adaptation to the case 
of persistence diagrams or other topological
structures (such as contour trees and Reeb graphs)
may enable an additional structural regularization,
this ought to be investigated in future work.
}


\section{Conclusion}
In this paper, we presented an original framework for tracking topological
features in a robust and efficient way. It is the first approach combining
topological data analysis and transport for feature tracking.
As the kernel of our approach,
we proposed a sparse-compliant extension of the seminal
assignment algorithm for the exact \maxime{matching} of persistence
diagrams, leveraging in practice important speedups.
We introduced a new metric for persistence diagrams that
enhances geometrical stability and further improves computation time.
Overall, in comparison \maxime{with} overlap-based techniques, our approach
displays improved performance and robustness to temporal
downsampling, as experiments have shown.

\maxime{We plan to release the implementation
of our tracking framework open-source as a part of TTK \cite{ttk}
in the near future;
we hope that it will be useful to the community with an interest for 
efficient tracking methods.}
We look forward to adapting \maxime{it to} tracking phenomena
in \emph{in-situ} contexts, where the large-scale
time-varying data is accessed in a streaming fashion.
\maxime{As we are also interested in larger \mbox{datasets},
we are currently carrying out scaling tests on complex physical case studies
available at Total S.A.,
for which one needs specifically adapted
rendering techniques \cite{Lukasczyk17} to apprehend the resulting graphical
complexity of the topology evolution.}

We also believe that the application potential of our matching framework
can be studied for tasks other than time-tracking,
for instance, self-pattern matching and symmetry detection \cite{thomas14},
or feature comparisons in ensemble data.



\maxime{
\acknowledgments{
\small{
This work is partially supported by the Bpifrance grant ``AVIDO'' (Programme
d'Investissements d'Avenir, reference P112017-2661376/DOS0021427) and by the
French National Association for Research and Technology (ANRT), in the framework
of the LIP6 - Total SA CIFRE partnership reference 2016/0010.
The authors would like to thank the anonymous reviewers for their thoughtful
remarks and suggestions.}
}}

\clearpage

\bibliographystyle{abbrv-doi}

\bibliography{template}
\end{document}